\UseRawInputEncoding
\pdfoutput=1
\documentclass[journal]{IEEEtran}   
\usepackage{multirow}
\usepackage{diagbox}
\usepackage{amssymb}
\usepackage{amsmath}
\usepackage{graphicx}
\usepackage{cite}
\usepackage{citesort}
\usepackage{booktabs}
\usepackage{subfigure}
\usepackage{graphicx,epstopdf}
\usepackage{epsfig}	
\usepackage{cite,graphicx,amsmath,amssymb}
\usepackage{comment}
\usepackage{amssymb}
\usepackage{amsmath}
\usepackage{cite}
\usepackage{url}
\usepackage{xcolor}
\usepackage{cite,graphicx,amsmath,amssymb}
\usepackage{subfigure}
\usepackage{citesort}
\usepackage{fancyhdr}
\usepackage{mdwmath}
\usepackage{mdwtab}
\usepackage{caption}
\usepackage{amsthm}
\usepackage{lipsum}
\usepackage{array}

\newtheorem{theorem}{Theorem}

\newtheorem{lemma}{Lemma}

\newtheorem{corollary}{Corollary}

\newtheorem{remark}{Remark}  

\makeatletter
\def\ScaleIfNeeded{%
\ifdim\Gin@nat@width>\linewidth \linewidth \else \Gin@nat@width
\fi } \makeatother

\begin{document}

\title{\Huge{Secrecy Performance Analysis of Multi-Functional RIS-Assisted NOMA Networks}}

\author{Yingjie~Pei,~\IEEEmembership{Graduate Student Member,~IEEE}, Wanli~Ni,~\IEEEmembership{Member,~IEEE}, Jin~Xu,~\IEEEmembership{Member,~IEEE}, Xinwei Yue,~\IEEEmembership{Senior Member,~IEEE}, Xiaofeng~Tao,~\IEEEmembership{Senior Member,~IEEE}, and Dusit Niyato,~\IEEEmembership{Fellow, IEEE}

\thanks{This work was supported by National Natural Science Foundation of China under Grant 61932005 and in part by the Fundamental Research Funds for the Central Universities under Grant 2242022k60006. \emph{(Corresponding author: Xiaofeng Tao.)}}
\thanks{Y. Pei, J. Xu and X. Tao are with the School of Information and Communication Engineering, Beijing University of Posts and Telecommunications, Beijing 100876, China (email: yingjie.pei@bupt.edu.cn, jin.xu@bupt.edu.cn, taoxf@bupt.edu.cn).}
\thanks{W. Ni is with the Department of Electronic Engineering, Tsinghua University, Beijing 100084, China, and also with the Beijing National Research Center for Information Science and Technology, Beijing 100084, China (e-mail: niwanli@tsinghua.edu.cn).}
\thanks{X. Yue is with the Key Laboratory of Information and Communication Systems, Ministry of Information Industry and also with the Key Laboratory of Modern Measurement $\&$ Control Technology, Ministry of Education, Beijing Information Science and Technology University, Beijing 102206, China (email: xinwei.yue@bistu.edu.cn).}
\thanks{D. Niyato is with the College of Computing and Data Science, Nanyang Technological University, Singapore 639798 (email: dniyato@ntu.edu.sg).}
}


\maketitle

\begin{abstract}
Although reconfigurable intelligent surface (RIS) can improve the secrecy communication performance of wireless users, it still faces challenges such as limited coverage and double-fading effect. To address these issues, in this paper, we utilize a novel multi-functional RIS (MF-RIS) to enhance the secrecy performance of wireless users, and investigate the physical layer secrecy problem in non-orthogonal multiple access (NOMA) networks. Specifically, we derive the secrecy outage probability (SOP) and secrecy throughput expressions of users in MF-RIS-assisted NOMA networks with external and internal eavesdroppers. The asymptotic expressions for SOP and secrecy diversity order are also analyzed under high signal-to-noise ratio (SNR) conditions. Additionally, we examine the impact of receiver hardware limitations and error transmission-induced imperfect successive interference cancellation (SIC) on the secrecy performance. Numerical results 
indicate that: i) under the same power budget, the secrecy performance achieved by MF-RIS significantly outperforms active RIS and simultaneously transmitting and reflecting RIS; ii) with increasing power budget, residual interference caused by imperfect SIC surpasses thermal noise as the primary factor affecting secrecy capacity; and iii) deploying additional elements at the MF-RIS brings significant secrecy enhancements for the external eavesdropping scenario, in contrast to the internal eavesdropping case.
\end{abstract}
\section{Introduction}
Future sixth-generation (6G) networks need to satisfy the demands for greater bandwidth, higher speeds, and lower latency \cite{Zhengquan20196G},\cite{you2021towards}. This necessitates trends towards higher radio frequencies and larger antenna arrays in 6G networks \cite{Saad2020vision6G}. 
Reconfigurable intelligent surface (RIS), as a two-dimensional implementation of electromagnetic metamaterials, intelligently regulate spatial electromagnetic waves with controllable amplitude, phase, and frequency in a programmable manner \cite{2020QingqingMag},\cite{ShiminGong2020RIS}. RIS has emerged as one of the highly anticipated candidate technologies for 6G in recent years \cite{2021QingqingRISTutorial}. Prior studies have demonstrated that deploying RIS can significantly enhance the coverage and capacity of wireless communication networks \cite{2020BeixiongRISOFDM},\cite{LiangYang2020RISCoverage}, 
enable facilitate sensing and localization \cite{Jingzhi2020RISSensing},\cite{Xinyi2022RISSensing}, and guarantee secure communication at the physical layer \cite{GuenSun2021RISUAVPLS},\cite{Yingjie2023RISPLS},\cite{Waqas2024RISPLS}.

One of the main limitations of traditional RIS is its ability to serve users only in a half-space \cite{Linglong2020RISPrototyping}. To overcome this issue, 
the simultaneously transmitting and reflecting RIS (STAR-RIS) with dual functionality was proposed in \cite{Yuanwei2021STAR}. Generally, STAR-RIS can control the magnitude and phase of incoming signals by adjusting the bias voltage at the embedded PIN diode, enabling signal propagation on both sides of the surface \cite{JiaqiXu2021STAR}. The authors in \cite{ChaoZhang2022STARNOMA} proposed three typical operating modes of STAR-RIS, namely energy splitting (ES), mode switching, and time switching. The authors of \cite{Papazafeiropoulos2022STARMIMO} studied the coverage characteristics of STAR-RIS-assisted massive antenna networks. The ergodic rate performance of STAR-RIS-aided full-duplex communication networks was analyzed and optimized in \cite{salem2023star}. By redesigning the transmission and reflection coefficients, STAR-RIS can provide communication networks with superior secrecy rate compared to RIS schemes \cite{Hehao2021STARPLS}. Furthermore, an active monitoring system assisted by STAR-RIS was proposed in \cite{Guojie2023STARActiveEve}, where STAR-RIS can enhance the eavesdropping performance of the monitor while ensuring its concealment.

Additionally, the excellent adaptability of STAR-RIS allows it to integrate with emerging multiple access technologies, such as non-orthogonal multiple access (NOMA), achieving mutual benefits \cite{Manzoor2023STAR}. The core idea of the NOMA technology is to differentiate users in the power domain to achieve ultra-high spectrum efficiency \cite{2022XinyueNGMA}. The authors in \cite{Boqun2022STARNOMA} studied the rate performance STAR-RIS-assisted NOMA networks, deriving closed-form and approximate expressions for ergodic rate. Furthermore, authors in \cite{Xinwei2023STARRIS} analyzed the outage behavior and throughput of users in STAR-RIS-assisted NOMA networks. The error characteristics of NOMA networks were explored via STAR-RIS in \cite{Farjam2023STARNOMA}, deriving closed-form solutions for block error rate and goodput. The authors in \cite{Brian2023SERSTARNOMA} proposed a scaling and rotating constellations method to match the optimal parameters of STAR-RIS, thereby improving the error-rate performance of uplink NOMA networks.

Although the collaboration between STAR-RIS and NOMA networks holds a prospective future, the full-space coverage transmission facilitated by STAR-RIS in NOMA networks exacerbates the risk of eavesdropping on users' private signals \cite{lv2024safeguarding}. For this issue, the secrecy performance of STAR-RIS-assisted NOMA networks was investigated in \cite{Xiangbin2023STARNOMAPLS}, where both closed-form and approximate expressions of secrecy rate were derived. The authors of \cite{li2022enhancing} considered the impact of residual hardware impairments at the transceiver on the outage behavior of private signals in STAR-RIS-secured NOMA networks. The secrecy rate of NOMA networks was maximized in \cite{YiHan2022STARNOMAPLS} with the assistance of artificial noise at the base station (BS) and the beamforming design at STAR-RIS.  
The aforementioned literatures only consider the external eavesdropping case. However, users located within the service area have the opportunity to wiretap others' information, making them internal Eves (I-Eves). The authors in \cite{HuiHan2023STARNOMAPLS} proposed a STAR-RIS-secured NOMA networks with the presence of an I-Eve, where the coefficients at STAR-RIS are adjusted to maximize the secure rate of users. Building upon this, the secrecy capacity of uplink NOMA networks was optimized in \cite{ZhengZhang2022STARPLS} with discrete phase-shifted STAR-RIS. Furthermore, the authors in \cite{YanboZhang2023STARNOMAPLS} enhanced the capability of STAR-RIS in guaranteeing the secrecy rate and outage behavior of NOMA networks by devising power allocation and pairing strategies under both external and internal eavesdropping scenarios.

These studies of STAR-RIS primarily focused on passive signal manipulation without accounting for the potential introduced by incorporating active components. Hence, a constraint on the deployment of STAR-RIS networks is the signal degradation due to the existence of multiplicative fading \cite{zhang2021active}. In response to the issue, a multi-functional RIS (MF-RIS) was recently proposed in \cite{WenWang2023MFRIS}, embedding with low-cost and power-controllable amplifiers to ensure signal propagation in full space while mitigating the influence of multiplicative fading \cite{Wanli2024MFRIS}. The outage behaviors and ergodic rate of MF-RIS over STAR-RIS were evaluated in \cite{Yue2024ASTRSNOMA}, where MF-RIS exhibits a lower outage probability with a small number of elements.
The authors in \cite{Ailing2023MFRIS} validated that by redesigning the reflection and refraction parameters, the throughput of MF-RIS networks exceeds both STAR-RIS and traditional RIS. The sum-rate optimization of MF-RIS-enhanced uplink networks was investigated in \cite{YingjieYan2023MFRIS} by redesigning the transmitting power and the received beamforming vector. The authors in \cite{Ailing2023NextGenMFRIS} comprehensively studied the sum-rate performance of MF-RIS-assisted NOMA networks. 
As a further advance, the authors in \cite{Anastasios2024ASTAR} surveyed the potential of MF-RIS in enhancing the sum-rate performance of massive multiple-input multiple-output networks over correlated fading channels.

\subsection{Motivations}
Although the aforementioned literature highlights the advantages of MF-RIS and NOMA cooperative communication networks in terms of reliable transmission and high throughput, the security issues at the physical layer inherent to MF-RIS cannot be overlooked. On one hand, the full-space radiation and signal amplification by MF-RIS make it easier for Eves to intercept the legitimate signals, with I-Eves potentially benefiting from the channel diversity gain provided by MF-RIS beamforming \cite{Hehao2021STARPLS},\cite{ZheZhang2021PLSRIS}. On the other hand, the impact of thermal noise generated by active components in MF-RIS on secure signal transmission remains unknown, which has never been considered in traditional RIS and STAR-RIS assisted networks \cite{Waqas2024RISPLS},\cite{salem2023star},\cite{zhang2021active}. The work in \cite{Xiangbin2023STARNOMAPLS} analyzed the secrecy performance of STAR-RIS-assisted NOMA networks but only considered the external eavesdropping scenario. In addition, the work in {\cite{YiHan2022STARNOMAPLS}}{-\cite{YanboZhang2023STARNOMAPLS}} surveyed the secrecy characteristics of STAR-RIS-assisted NOMA networks from an optimization perspective, lacking theoretical analysis. In this paper, we aim to explore the potential capability of MF-RIS in safeguarding NOMA networks by seeking answers to these questions.

\begin{itemize}
  \item How does the significant thermal noise generated by the active components in MF-RIS affect the network's secrecy communications?
  \item When comparing imperfect successive interference cancellation (SIC) with thermal noise, which factor poses a greater risk to the secrecy performance of MF-RIS-assisted NOMA networks?
  \item What are the disparities in the secrecy performance of MF-RIS-assisted NOMA networks across various eavesdropping scenarios?
\end{itemize}

\subsection{Contributions}
To the best of our knowledge, this paper presents the first comprehensive analysis of the physical layer secrecy performance of MF-RIS-assisted NOMA networks. Specifically, we delve into the secure communication capabilities of randomly distributed users within these networks, taking into account both external and internal eavesdropping threats. Additionally, we derive approximate and asymptotic expressions of the secrecy outage probability (SOP) and secrecy throughput of users\footnote{Although the derived expressions for SOP and secrecy throughput are approximate, they provide valuable insights into system design. Specifically, these expressions allow us to quantitatively evaluate how key factors, such as the number of MF-RIS elements, total power budget, and hardware limitations, impact the secure communication performance of NOMA networks.}. For the purpose of being more realistic, the residential interference and thermal noise caused by hardware limitations or electron motion are also taken into account. Furthermore, we also evaluate the impact total power budget and the number of reconfigurable elements on the secure communication. The major contributions of this paper can be summarized as follows:
\begin{enumerate}
  \item We propose an MF-RIS-secured NOMA network, leveraging the MF-RIS to enhance the quality of legitimate cascaded channels and improve the secrecy capability of users through coherent phase shifting. To extensively evaluate the secrecy performance, we consider both external and internal wiretapping scenarios. Additionally, we select the SOP of randomly distributed users as the key metric to demonstrate the superior secrecy capabilities of MF-RIS compared to STAR-RIS and active RIS schemes.
  \item For the external wiretapping scenarios, an illegal external Eve (E-Eve) attempts to intercept the private messages of users served by MF-RIS. We firstly examine the statistical property of the wiretapped cascaded channels. Taking into account hardware limitations at receivers, we then derive approximate expressions of the SOP and secrecy throughput of both users, considering both imperfect and perfect SIC cases. To gain further insights, we also obtain asymptotic SOP expressions as the transmit power approaches infinity. On this basis, the secrecy diversity order is attained and proportional to the number of MF-RIS elements.
  \item For the internal wiretapping scenarios, the user positioned within the reflection region is designated as the I-Eve. The I-Eve is capable of decoding the private information transmitted to the user on the opposite side. Given that the I-Eve has been registered in the serving cell, it can directly benefit from the channel diversity gain provided by MF-RIS beamforming. To address this challenge, we re-examine the characteristics of the wiretapping channels and derive the SOP expressions of the user located in the refraction region under both imperfect and perfect SIC cases. However, due to the presence of imperfect SIC and the I-Eve, the secrecy diversity order of the other users on the opposite side consistently remains at zero.
  \item Numerical results validate the accuracy of our theoretical analyses and highlight several key findings: Firstly, within the same power budget, MF-RIS-secured NOMA networks outperform both STAR-RIS and active RIS-based schemes in terms of SOP and secrecy throughput. Secondly, while increasing the number of MF-RIS elements significantly improves user secrecy in external eavesdropping scenarios, its impact may be limited in internal eavesdropping scenarios due to the I-Eve's ability to leverage the same diversity gain. Finally, both thermal noise at the MF-RIS and residual interference from imperfect SIC impact private signal transmission, with the latter becoming the predominant factor as the power budget increases.
\end{enumerate}


\section{System Model}\label{SectionII}
\subsection{Network Descriptions}
We consider an MF-RIS-secured NOMA network as illustrated in Fig. 1, which contains a BS, an MF-RIS, randomly distributed users and a malicious E-Eve\footnote{Note that the proposed system model only considers one single Eve in both external and internal eavesdropping scenarios. Designing the secrecy transmission of an MF-RIS-assisted NOMA network under non-cooperative and cooperative wiretapping attacks with multiple Eves are more complex and challenging \cite{WenWang2023UAVRISPLS},\cite{WenWang2022STARPLS}, which will be considered in our future work.
}. Due to the existence of obstacles and sever attenuation, the direct links from the BS to users and the E-Eve are assumed to be totally blocked and communication links can only be established via the assistance of MF-RIS. To provide straightforward analyses, each node is equipped with a single antenna\footnote{Given that this paper primarily focuses on the physical layer security enhancements provided by MF-RIS in NOMA networks, we simplified the architecture of BS and assumed it is equipped with a single antenna \cite{Farjam2023STARNOMA},\cite{LuLv2022CovertRISNOMA}. This assumption can be extended to multi-antenna scenarios by considering beamforming vector and updating the corresponding channel matrices.}. The MF-RIS is composed of \emph{M}  reconfigurable elements embedded with active loads like delay lines and power amplifier \cite{WenWang2023MFRIS}. For the sake of efficient resource utilization and high diversity gain, the MF-RIS operates under ES working mode. The amplitude coefficients for reflection and refraction are denoted as ${e_r}$ and ${e_t}$, respectively, which satisfies the relationship ${e_r} + {e_t} = 1$ based on the conservation of energy \cite{Yuanwei2021STAR}. 
As a consequence, the superimposed signal is emitted from the BS to MF-RIS firstly, and then transmitted to reflection and refraction users simultaneously.
\begin{figure}[t!]
    \begin{center}
        \includegraphics[width=2.8in,  height=2.0in]{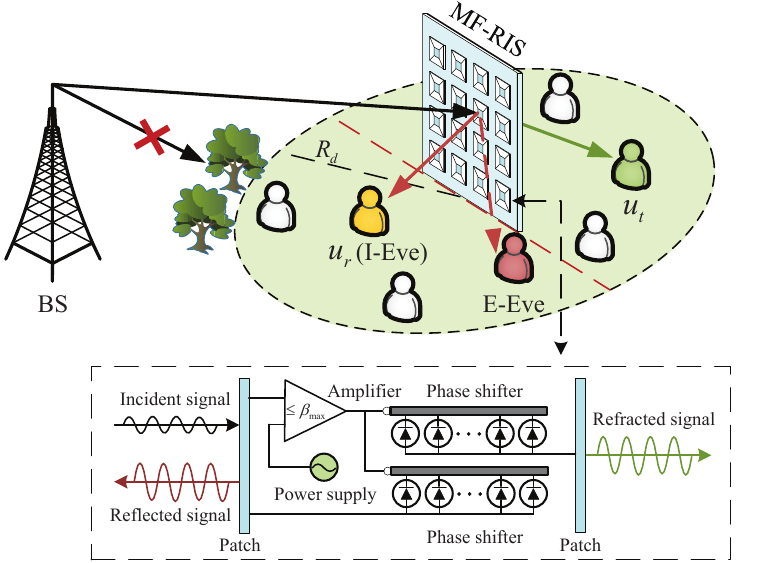}
        \caption{An illustration of MF-RIS-secured NOMA networks.}
        \label{system_model}
    \end{center}
\end{figure}
In the considered MF-RIS-secured NOMA network, there exists a line-of-sight (LoS) link between the BS and MF-RIS \cite{YiHan2022STARNOMAPLS}. The users are randomly located within a ring of radius $R_d$. Users located in the reflection space of the MF-RIS are termed as reflection users ($u_r$), while those situated on the opposite side are regarded as refraction users ($u_t$). Following the NOMA principle, we initially select one user, each from the reflection and refraction regions to form NOMA pairs. Subsequently, their signals are combined on the same subcarrier for non-orthogonal transmission with the assistance of MF-RIS, and at the receiver, the superimposed signals are extracted using SIC technology at the power domain level. Note that, in practical implementation, the radius of the MF-RIS service area is usually not large for establishing a stable LoS link and the path loss impact from the MF-RIS to users has little difference. Hence, the order of signal decoding when utilizing SIC technology primarily depends on the values of $e_r$ and $e_t$ \cite{ChaoZhang2022STARNOMA}. The distances of $u_r$ and $u_t$ from the MF-RIS position are represented as $d_{r}$ and $d_{t}$, respectively, and their probability density functions are given as

\begin{small}
\begin{align}\label{pdf drr drt}
{f_{{d_{\varphi }}}}\left( x \right) = \frac{2}{{\pi R_d^2}} \cdot \frac{\partial }{{\partial x}}\int_0^\pi  {\int_0^x r } drd\theta  = \frac{{2x}}{{R_d^2}},
\end{align} where $\varphi  \in \left\{ {r,t} \right\}$.
\end{small}

\subsection{Channel Characteristics}
The links from BS to MF-RIS and from MF-RIS to $u_\varphi$ and E-Eve are denoted as ${{\mathbf{H}}_{b}}$, ${{\mathbf{H}}_{\varphi}}$, and ${{\mathbf{H}}_{e}}$, respectively. The complex channel gains are given as
\begin{small}
\begin{align}
{{\mathbf{H}}_{\eta}} = \sqrt {\chi d_{\eta}^{ - \alpha }} {{\mathbf{h}}_{\eta}} = \sqrt {\chi d_{\eta}^{ - \alpha }} {\left[ {{h_{\eta ,1}}, \cdots ,{h_{\eta ,m}}, \cdots ,{h_{\eta ,M}}} \right]^H},
\end{align}
\end{small}where ${{\mathbf{H}}_{b}}$ and ${{\mathbf{H}}_{\varphi}}$ are supposed to be Rician fading channels, $h_{\eta,m} = {\sqrt {\frac{\kappa_{\eta} }{{\kappa_{\eta}  + 1}}}  + \sqrt {\frac{1}{{\kappa_{\eta}  + 1}}} \tilde h_{\eta,m}} $, $\tilde h_{\eta,m} \sim \mathcal{C}\mathcal{N}\left( {0,1} \right)$, $\eta \in \left\{ {b,r,t,e} \right\}$, and $\varphi \in \left\{ {r,t} \right\}$. The frequency dependent factor is denoted as $\chi $. The terms $d_b$ and $d_e$ represent the distance from BS to MF-RIS and from MF-RIS to E-Eve, respectively. The term $\alpha $ is the path loss exponent. The Rician factor is given by $\kappa_{\eta}$ and the Rician fading degenerates into Rayleigh fading by setting $\kappa_{\eta} = 0$.
Since the malicious E-Eve is unregistered in the cell, it cannot be served by the pre-designed beamforming of MF-RIS and benefit from the channel diversity gains, which results in almost no strong LoS components in the signal received at E-Eve. Therefore, the links between MF-RIS and E-Eve are modeled as Rayleigh fading channels \cite{YiHan2022STARNOMAPLS},\cite{ZhengZhang2023STARPLS}. We suppose that the perfect channel state information (CSI) of users can be obtained at BS by means of feedback or training processes, except for the E-Eve's since it hardly interacts with other nodes and often remains silent to avoid detection \cite{Yingjie2023ARISNOMA}.

\subsection{Signal Model}
According to NOMA protocol, the superimposed signal is transmitted from BS to MF-RIS and then sent to users simultaneously. The signal received at ${u_\varphi }$ can be given by
\begin{small}
\begin{align}\label{the received signal at users}
{y_\varphi } = {\mathbf{H}}_{\varphi }^H{{\mathbf{\Theta }}_\varphi }{{\mathbf{H}}_{b}}\sum\limits_{{\bar \varphi }  \in \left\{ {r,t} \right\}} {\sqrt {{P_b}{a_{\bar \varphi } }} {x_{\bar \varphi } }}  + {\mathbf{H}}_{\varphi }^H{{\mathbf{\Theta }}_\varphi }{{\mathbf{n}}_s} + n,
\end{align}
\end{small}where ${P_{b}}$ denotes the transmit power at the BS and ${{x_\varphi}}$ is the confidential signal of $u_{\varphi}$ with unity power, i.e., $\mathbb{E}\{ {\left| {{x_\varphi}} \right|^2}\}  = 1$, ${\varphi  \in \left\{ {r,t} \right\}}$. To ensure user fairness, ${{a_\varphi}}$ represents the power allocation factor of $u_\varphi$, which satisfies the relationship ${a_r} + {a_t} = 1$. In this paper, we suppose that $u_r$ is the strong user with superior channel condition compared with $u_t$, and thus ${a_t} > {a_r}$. The diagonal matrix for $u_\varphi$ is represented by ${{\mathbf{\Theta }}_\varphi } = diag\left[ {\sqrt {\beta _{\varphi,1}} {e^{j\theta _{\varphi,1}}}, \cdots ,\sqrt {\beta _{\varphi,m}} {e^{j\theta _{\varphi,m}}}, \cdots ,\sqrt {\beta _{\varphi,M}} {e^{j\theta _{\varphi,M}}}} \right]$, where $\beta _{\varphi,m}$ indicates the amplification factor for $u_\varphi$ of the \emph{m}-th MF-RIS elements, ${\beta _{\varphi,m}} \in \left[ {0,{\beta _{\max }}} \right]$ and ${\beta _{\max }} \geqslant 1$. In order to reduce the complexity of phase shift configuration, it is assumed that $\beta _{\varphi,1} = \beta _{\varphi,2} =  \cdots  = \beta _{\varphi,M} = \beta _\varphi$, ${\varphi  \in \left\{ {r,t} \right\}}$ and $m = 1,2, \cdots ,M$. In addition, ${{\mathbf{n}}_s} \sim \mathcal{C}\mathcal{N}\left( {{\mathbf{0}},\sigma _s^2{{\mathbf{1}}_M}} \right)$ denotes the thermal noise caused by the active components at MF-RIS like asymmetric current mirrors and current-inverting converters, where ${{\mathbf{1}}_M} \in {\mathbb{C}^{M \times 1}}$ indicates the all-ones column vector \cite{2022LinglongSunARIS}. The $n \sim \mathcal{C}\mathcal{N}\left( {0,\sigma _n^2} \right)$ is the additive white
Gaussian noise (AWGN) at the users with mean power parameter $\sigma _n^2$.

Different from the passive RIS and STAR-RIS, the MF-RIS embeds a dedicated amplifier in each element to provide additional power gain to the incident signals, thereby further enhancing the reliability of LoS links. In this case, the total power consumed by MF-RIS-assisted NOMA networks can be represented as ${P_{tot}} = {P_b} + M{P_r} + 2M\left( {{P_{ps}} + {P_{dc}}} \right)$, where $P_b$ and $P_r$ represent the transmit power at BS and the amplifier power at each element, ${P_{ps}}$ and ${P_{dc}}$ denote the power consumed by phase shifters and direct current biasing circuits \cite{2022CunhuaPanARIS}, respectively.

Recall that the SIC decoding order mainly depends on the ES coefficients and $u_r$ is supposed to be the strong user, we assume more energy is assigned to reflection region, i.e., ${e_r} \geqslant {e_t}$, which means $u_r$ has better cascaded channel conditions than $u_t$. Accordingly, the SINR for $u_r$ to decode $u_t$'s information can be given by
\begin{small}
\begin{align}\label{SINR r decode t}
{\gamma _{r,t}} = \frac{{{P_b}{a_t}{\beta _r}{{\left| {{\mathbf{H}}_{r}^H{{\mathbf{\Phi }}_r}{{\mathbf{H}}_{b}}} \right|}^2}}}{{{P_b}{a_r}{\beta _r}{{\left| {{\mathbf{H}}_{r}^H{{\mathbf{\Phi }}_r}{{\mathbf{H}}_{b}}} \right|}^2} + {\beta _r}{{\left| {{\mathbf{H}}_{r}^H{{\mathbf{\Phi }}_r}{\mathbf{n}}_{s}} \right|}^2} + \sigma _n^2}},
\end{align}
\end{small}where ${{\mathbf{\Phi }}_\varphi } = diag\left[ {{e^{j\theta _{\varphi,1}}}, \cdots ,{e^{j\theta _{\varphi,m}}}, \cdots ,{e^{j\theta _{\varphi,M}}}} \right]$. After removing $u_t$'s message, the SINR for $u_r$ to decode its own information with imperfect SIC is shown as follows
\begin{small}
\begin{align}\label{SINR r decode r ipSIC}
\gamma _{r,r}^{ipSIC} = \frac{{{P_b}{a_r}{\beta _r}{{\left| {{\mathbf{H}}_{r}^H{{\mathbf{\Phi }}_r}{{\mathbf{H}}_{b}}} \right|}^2}}}{{{\beta _r}{{\left| {{\mathbf{H}}_{r}^H{{\mathbf{\Phi }}_r}{\mathbf{n}}_{s}} \right|}^2} + \varpi {P_b}{{\left| {h_r^{ip}} \right|}^2} + \sigma _n^2}},
\end{align}
\end{small}where $\varpi  \in \left[ {0,1} \right]$ denotes the residual interference level for SIC. Specifically, $\varpi  = 0$ and $\varpi  \ne 0$ represent the switchover between perfect and imperfect SIC, respectively. Without loss of generality, the residual interference at $u_\varphi$ generated by imperfect SIC is modeled as the Rayleigh random variable and the relative complex channel parameter is denoted by ${h_\varphi^{ip}} \sim \mathcal{C}\mathcal{N}\left( {0,{\Omega_\varphi^{ip}}} \right)$, ${\varphi  \in \left\{ {r,t} \right\}}$ \cite{Xinwei2023STARRIS}.

Unlike the $u_r$ who has superior channel condition, $u_t$  will be allocated with a larger power coefficient by BS since it has inferior LoS link qualities, i.e., ${a_t} > {a_r}$. Thus, $u_t$ can decode its own information directly from the superimposed signal without the assistance of SIC and the corresponding SINR can be given as

\begin{small}
\begin{align}\label{SINR t decode t}
{\gamma _{t,t}} = \frac{{{P_b}{a_t}{\beta _t}{{\left| {{\mathbf{H}}_{t}^H{{\mathbf{\Phi }}_t}{{\mathbf{H}}_{b}}} \right|}^2}}}{{{P_b}{a_r}{\beta _t}{{\left| {{\mathbf{H}}_{t}^H{{\mathbf{\Phi }}_t}{{\mathbf{H}}_{b}}} \right|}^2} + {\beta _t}{{\left| {{\mathbf{H}}_{t}^H{{\mathbf{\Phi }}_t}{\mathbf{n}}_{s}} \right|}^2} + \sigma _n^2}}.
\end{align}
\end{small}
\subsection{Wiretapping Scenarios}
Due to the open nature of wireless channels, signals are prone to interception by malicious Eves, leading to the leakage of users' private information. In order to further investigate the physical layer secrecy performance of the MF-RIS-assisted NOMA networks, we take into account two typical wiretapping cases, i.e., external and internal eavesdropping scenarios.

\subsubsection{External Wiretapping Scenario}
In the external wiretapping scenario, the Eve is not a legitimate registered user in the network, but can still receive the superimposed signals from the MF-RIS. In this paper, we deploy the E-Eve in the reflection area, which presents a challenging situation because more energy is split to the reflection signals by MF-RIS. The received signal at E-Eve is given by

\begin{small}
\begin{align}\label{received signal at E-Eve}
{y_{ee}} = {\mathbf{H}}_{e}^H{{\mathbf{\Theta }}_r}{{\mathbf{H}}_{b}}\sum\limits_{{\bar \varphi }  \in \left\{ {r,t} \right\}} {\sqrt {{P_b}{a_{\bar \varphi } }} {x_{\bar \varphi } }}  + {\mathbf{H}}_{e}^H{{\mathbf{\Theta }}_r}{{\mathbf{n}}_s} + {n_e},
\end{align}
\end{small}where ${n_e} \sim \mathcal{C}\mathcal{N}\left( {0,\sigma _e^2} \right)$ indicates the AWGN at E-Eve with mean power parameter $\sigma _e^2$.

Furthermore, the SINR for E-Eve to intercept $u_t$'s information can be given by
\begin{small}
\begin{align}\label{SINR EE decode t}
{\gamma _{ee,t}} = \frac{{{P_b}{a_t}{\beta _r}{{\left| {{\mathbf{H}}_{e}^H{{\mathbf{\Phi }}_r}{{\mathbf{H}}_{b}}} \right|}^2}}}{{{P_b}{a_r}{\beta _r}{{\left| {{\mathbf{H}}_{e}^H{{\mathbf{\Phi }}_r}{{\mathbf{H}}_{b}}} \right|}^2} + {\beta _r}{{\left| {{\mathbf{H}}_{e}^H{{\mathbf{\Phi }}_r}{\mathbf{n}}_{s}} \right|}^2} + \sigma _e^2}}.
\end{align}
\end{small}Similarly, the SINR for E-Eve to intercept $u_r$'s information with imperfect SIC can be given by

\begin{small}
\begin{align}\label{SINR EE decode r ipsic}
{\gamma _{ee,r}^{ipSIC}} = \frac{{{P_b}{a_r}{\beta _r}{{\left| {{\mathbf{H}}_{e}^H{{\mathbf{\Phi }}_r}{{\mathbf{H}}_{b}}} \right|}^2}}}{{{\beta _r}{{\left| {{\mathbf{H}}_{e}^H{{\mathbf{\Phi }}_r}{\mathbf{n}}_{s}} \right|}^2} + \varpi {P_b}{{\left| {{h_{e,r}^{ip}}} \right|}^2} + \sigma _e^2}},
\end{align}
\end{small}where ${h_{e,r}^{ip}} \sim \mathcal{C}\mathcal{N}\left( {0,{\Omega_{e,r}^{ip}}} \right)$ indicates the impact of residential interference brought by imperfect SIC on E-Eve.

\subsubsection{Internal Wiretapping Scenario}
Unlike the aforementioned external eavesdropping scenario, the I-Eve is a registered user within the NOMA service cell and can benefit from the cascaded channel gains brought by the MF-RIS \cite{NaLi2021IRSNOMAPLS}. For this scenario, we still consider a worst-case situation where $u_r$ is identified as the I-Eve. Note that, to protect the signal privacy of $u_t$ in the internal wiretapping scenario, the BS will adjust the power allocation of $u_t$, making ${a_t} < {a_r}$, to ensure that its signal is decoded last during SIC at the receiver. Additionally, MF-RIS will also assign more energy to the refraction region to increase the secrecy capacity of $u_t$. As a consequence, the SINR for $u_t$ to decode its own information and I-Eve to intercept $u_t$'s information with imperfect SIC can be given by

\begin{small}
\begin{align}\label{t decode t ipsic IE}
\gamma _{t,t}^{ipSIC} = \frac{{{P_b}{a_t}{\beta _t}{{\left| {{\mathbf{H}}_{t}^H{{\mathbf{\Phi }}_t}{{\mathbf{H}}_{b}}} \right|}^2}}}{{\varpi {P_b}{{\left| {{h_t^{ip}}} \right|}^2} + {\beta _t}\left| {{\mathbf{H}}_{t}^H{{\mathbf{\Phi }}_t}{\mathbf{n}}_{s}} \right| + \sigma _n^2}},
\end{align}
\end{small}and
\begin{small}
\begin{align}\label{IE decode t ipsic}
{\gamma _{ie,t}^{ipSIC}} = \frac{{{P_b}{a_t}{\beta _r}{{\left| {{\mathbf{H}}_{r}^H{{\mathbf{\Phi }}_r}{{\mathbf{H}}_{b}}} \right|}^2}}}{{{\beta _r}{{\left| {{\mathbf{H}}_{r}^H{{\mathbf{\Phi }}_r}{\mathbf{n}}_{s}} \right|}^2} + \varpi {P_b}{{\left| {{h_{r,t}^{ip}}} \right|}^2} + \sigma _e^2}},
\end{align}
\end{small}respectively, where ${h_{r,t}^{ip}} \sim \mathcal{C}\mathcal{N}\left( {0,{\Omega_{r,t}^{ip}}} \right)$ represents the residential interference at I-Eve.

\section{Statistical Properties for Channels}\label{SectionIII}
In this section, we investigate the secrecy performance of users in the MF-RIS-assisted NOMA networks. Specifically, we obtain the statistical characteristics of the legitimate and eavesdropping cascade channels to lay a foundation for the subsequent theoretical analysis of the user's SOP performance.

\subsection{Statistical Properties for Legitimate Channels}
Due to the independent phase-adjusting capability of each reconfigurable element, MF-RIS can attain coherent alignment between the incident channels and the reflection or refraction channels, leveraging user-specific CSI. As a consequence, the cascaded channels from BS to MF-RIS and then to $u_\varphi$ can be defined as
${{\mathbf{H}}_{b\varphi }} = {\mathbf{H}}_{\varphi }^H{{\mathbf{\Theta }}_\varphi }{{\mathbf{H}}_{b}} = {\Xi _{b\varphi }}\sum\nolimits_{m = 1}^M {\left| {h_{\varphi ,m}h_{b,m}} \right|}$, where the phase shifts at the \emph{m}-th element $\theta _{\varphi,m} = \angle h_{\varphi,m } - \angle h_{b,m}$, ${\Xi _{b\varphi }} = \chi \sqrt {d_{b}^{ - \alpha }d_{\varphi }^{ - \alpha }{\beta _\varphi }} $, and $\sum\nolimits_{m = 1}^M {\left| {h_{\varphi,m}h_{b,m}} \right|} $ can be further defined as ${{{{\mathbf{\hat H}}}_{b\varphi }}}$, ${\varphi  \in \left\{ {r,t} \right\}}$. This capability serves to significantly enhance the cascaded channel gains. For the convenience of subsequent derivations, we provide the following lemma to describe the statistical characteristics of legitimate channels.
\begin{lemma} \label{Lemma1}
Upon utilizing the coherent phase-shifting scheme, the CDF and PDF of ${\left| {{{{\mathbf{\hat H}}}_{b\varphi }}} \right|^2}$ for $u_\varphi$ can be given as
\begin{small}
\begin{align}\label{cdf user channel}
{F_{{{\left| {{{{\mathbf{\hat H}}}_{b\varphi }}} \right|}^2}}}\left( x \right) = \frac{1}{{\Gamma \left( {{k_\varphi }} \right)}}\gamma \left( {{k_\varphi },\frac{{\sqrt x }}{{{l_\varphi }}}} \right),
\end{align}
\end{small}
and
\begin{small}
\begin{align}\label{pdf user channel}
{f_{{{\left| {{{{\mathbf{\hat H}}}_{b\varphi }}} \right|}^2}}}\left( x \right) = \frac{{{x^{\frac{{{k_\varphi }}}{2} - 1}}}}{{2l_\varphi ^{{k_\varphi }}\Gamma \left( {{k_\varphi }} \right)}}{e^{ - \frac{{\sqrt x }}{{{l_\varphi }}}}},
\end{align}
\end{small}where $\gamma \left( {a,x} \right) = \int_0^x {{p^{a - 1}}{e^{ - p}}dp} $ indicates the lower incomplete Gamma function \emph{\cite[Eq. (8.350.1)]{gradvstejn2000table}} and $\Gamma \left( x \right) = \int_0^\infty  {{e^{ - p}}{p^{x - 1}}dp} $ represents the Gamma function \emph{\cite[Eq. (8.310.1)]{gradvstejn2000table}}. In addition, ${k_\varphi } = M{\left[ {\mathbb{E}\left( {\left| {h_{\varphi,m}h_{b,m}} \right|} \right)} \right]^2}/\mathbb{D}\left( {\left| {h_{\varphi,m}h_{b,m}} \right|} \right)$ and ${l_\varphi } = \mathbb{D}\left( {\left| {h_{\varphi,m}h_{b,m}} \right|} \right)/\mathbb{E}\left( {\left| {h_{\varphi,m}h_{b,m}} \right|} \right)$. The expectation and variance of single Rician cascaded channel are respectively given as
\begin{small}
\begin{align}\label{mean single Rician channel}
\mathbb{E}\left( {\left| {h_{\varphi,m}h_{b,m}} \right|} \right) = \frac{{ {}_1{F_1}\left( { - \frac{1}{2};1; - {\kappa _{b}}} \right){}_1{F_1}\left( { - \frac{1}{2};1; - {\kappa _{\varphi }}} \right)}}{{4{\pi^{-1}}\sqrt {\left( {{\kappa _{b}} + 1} \right)\left( {{\kappa _{\varphi }} + 1} \right)} }},
\end{align}
\end{small}
and
\begin{small}
\begin{align}\label{var single Rician channel}
\mathbb{D}\left( {\left| {h_{\varphi,m}h_{b,m}} \right|} \right) =& 1 - \frac{{{{\left( {{}_1{F_1}\left( { - \frac{1}{2};1; - {\kappa _{b}}} \right)} \right)}^2}}}{{16\left( {{\kappa _{b}} + 1} \right)\left( {{\kappa _{\varphi }} + 1} \right)}}\notag \\ &\times{\left( {\pi {}_1{F_1}\left( { - \frac{1}{2};1; - {\kappa _{\varphi }}} \right)} \right)^2},
\end{align}
\end{small}referring to \emph{\cite{simon2002probability}}. ${}_1{F_1}\left( {a;b;z} \right) = \sum\nolimits_{n = 0}^\infty  {} \left( {{a^{\left( n \right)}}{z^n}} \right)/\left( {{b^{\left( n \right)}}n!} \right)$ indicates the generalized hypergeometric series which is introduced from Kummer's function of the first kind \emph{\cite[Eq. (9.14.1)]{gradvstejn2000table}}.
\begin{proof}
Referring to \cite{primak2005stochastic}, the approximated PDF expression of ${{{{\mathbf{\hat H}}}_{b\varphi }}}$ can be derived by utilizing Laguerre expansion as
\begin{small}
\begin{align}\label{pdf brfai hat}
{f_{\left| {{{{\mathbf{\hat H}}}_{b\varphi }}} \right|}}\left( x \right) = \frac{{{x^{{k_\varphi } - 1}}}}{{{l_\varphi }^{{k_\varphi }}\Gamma \left( {{k_\varphi }} \right)}}{e^{ - \frac{x}{{{l_\varphi }}}}}.
\end{align}
\end{small}Furthermore, the expectation of ${{{{\mathbf{\hat H}}}_{b\varphi }}}$ is in accordance with $\mathbb{E}\left( {| {{{{\mathbf{\hat H}}}_{b\varphi }}} |} \right) = M\mathbb{E}\left( { {\left| {h_{\varphi,m}h_{b,m}} \right|} } \right)$ due to the linear characteristic of the expectation operation. Similar, since the cascaded Rician channels are assumed to be independent from each other, we can attain $\mathbb{D}\left( {| {{{{\mathbf{\hat H}}}_{b\varphi }}} |} \right) = M\mathbb{D}\left( { {\left| {h_{\varphi,m}h_{b,m}} \right|} } \right)$.
On this basis, the PDF of ${| {{{{\mathbf{\hat H}}}_{b\varphi }}} |^2}$ is obtained as in (\ref{pdf user channel}). With the assistance of integral operation, the CDF expression of ${| {{{{\mathbf{\hat H}}}_{b\varphi }}} |^2}$ is further derived as (\ref{cdf user channel}), which completes the proof.
\end{proof}
\end{lemma}

\subsection{Statistical Properties for Eavesdropping Channels}
Given that the E-Eve consistently maintains silence and avoids interacting with other network nodes, acquiring CSI for the eavesdropping channel becomes challenging for the BS. Consequently, we utilize statistical methods to evaluate the characteristics of the eavesdropping channels. Since the CSI of E-Eve is unavailable, we define that ${{\mathbf{H}}_{be}} = {\mathbf{H}}_{e}^H{{\mathbf{\Theta }}_r}{{\mathbf{H}}_{b}} = {\Xi _{be}}\sum\nolimits_{m = 1}^M {\left| {h_{e,m}h_{b,m}} \right|{e^{j\theta _{d,m}}}} $, where ${\Xi _{be}} = \chi \sqrt {d_{b}^{ - \alpha }d_{e}^{ - \alpha }{\beta _r}} $. Considering the phase configuration of MF-RIS that has been completed based on the user's CSI, the phase difference is indicated as $\theta _{d,m} = \angle h_{b,m} - \angle h_{e,m}$. In addition, the phases of $h_{b,m}$ and $h_{e,m}$ are also random variables following a uniform distribution, i.e., $\angle h_{b,m},\angle h_{e,m} \sim U\left( {0,2\pi } \right)$, $m = 1,2, \cdots ,M$. In this case, the PDF of  phase difference ${\theta _{d,m}}$ is given by

\begin{small}
\begin{align}\label{pdf phase diff 1}
{f_{{\theta _{d,m}}}}\left( x \right) = \frac{1}{{2\pi }}\left( {1 + \frac{x}{{2\pi }}} \right),x \in \left[ { - 2\pi ,0} \right],
\end{align}
\end{small}
and
\begin{small}
\begin{align}\label{pdf phase diff 2}
{f_{{\theta _{d,m}}}}\left( x \right) = \frac{1}{{2\pi }}\left( {1 - \frac{x}{{2\pi }}} \right),x \in \left( {0,2\pi } \right].
\end{align}
\end{small}

\begin{lemma} \label{Lemma2}
The E-Eve is unable to achieve coherent matching of its own channels using MF-RIS, and thus the phase difference of the wiretapping cascade channels cannot be eliminated. In such a scenario, the CDF of $|{{{\mathbf{H}}}_{be}}|^2$ is given by

\begin{small}
\begin{align}\label{cdf eve channel squ}
{F_{|{{{\mathbf{H}}}_{be}}{|^2}}}\left( x \right) = 1 - {e^{ - \upsilon x}},
\end{align} where $\upsilon  = 2/\left( {M{{\left( {{\Xi _{be}}} \right)}^2}} \right)$.
\end{small}

\begin{proof}
With the assistance of Euler's formula, the real and imaginary component of cascaded wiretapping channels are denoted as $\operatorname{Re} \left\{ {{{\mathbf{H}}_{be}}} \right\} = {\Xi _{be}}\sum\nolimits_{m = 1}^M {\left| {h_{e,m}h_{b,m}} \right|} \cos \theta _{d,m}$ and $\operatorname{Im} \left\{ {{{\mathbf{H}}_{be}}} \right\} = {\Xi _{be}}\sum\nolimits_{m = 1}^M {\left| {h_{e,m}h_{b,m}} \right|} \sin \theta _{d,m}$, respectively. Since $\cos \theta _{d,m}$ and $\left| {h_{e,m}h_{b,m}} \right|$ are both independent random variables, we can calculate the mean of $\cos \theta _{d,m}$ as follows:

\begin{small}
\begin{align}\label{mean cos phase diff}
\mathbb{E}\left( {\cos \theta _{d,m}} \right) = \int_{ - 2\pi }^0 {\cos \theta _{d,m}{f_{\theta _{d,m}}}\left( {\theta _{d,m}} \right)} d\theta _{d,m}.
\end{align}
\end{small}Upon substituting (\ref{pdf phase diff 1}) into (\ref{pdf phase diff 2}), we can obtain $\mathbb{E}\left( {\cos \theta _{d,m}} \right) = 0$. Similarly, we can further obtain

\begin{small}
\begin{align}
\mathbb{E}\left( {{{\cos }^2}\theta _{d,m}} \right) = \int_{ - 2\pi }^0 {{{\cos }^2}\theta _{d,m}{f_{\theta _{d,m}}}\left( {\theta _{d,m}} \right)} d\theta _{d,m} = \frac{1}{4}.
\end{align}
\end{small}Note that the identical conclusion can be reached for the case $\theta _{d,m} \in \left( {0,2\pi } \right]$. For the imaginary component, we have $\mathbb{E}\left( {\sin \theta _{d,m}} \right) = 0$. On this basis, we suppose that $X = {\Xi _{be}}\left| {h_{e,m}h_{b,m}} \right|$ and $Y = \cos \theta _{d,m}$. Hence, the variance value of ${\Xi _{be}}\left| {h_{e,m}h_{b,m}} \right|\cos \theta _{d,m}$ is given by

\begin{small}
\begin{align}\label{var XY}
\mathbb{D}\left( {XY} \right) &= \mathbb{D}\left( X \right)\mathbb{D}\left( Y \right) + \mathbb{D}\left( X \right){\mathbb{E}^2}\left( Y \right) + {\mathbb{E}^2}\left( X \right)\mathbb{D}\left( Y \right)\notag \\ &= {\left( {{\Xi _{be}}} \right)^2}/4,
\end{align}
\end{small}where ${\Xi _{be}} = \chi \sqrt {d_{b}^{ - \alpha }d_{e}^{ - \alpha }{\beta _r}} $. Based on CLT, both $\operatorname{Re} \left\{ {{{\mathbf{H}}_{be}}} \right\}$ and $\operatorname{Im} \left\{ {{{\mathbf{H}}_{be}}} \right\}$ follow Gaussian distributions with a mean of zero and a variance of $M{\left( {{\Xi _{be}}} \right)^2}/4$. As a consequence, the cascaded wiretapping channel ${{{\mathbf{H}}_{be}}}$ can be regarded as the zero mean circle symmetric complex gaussian random variable, i.e., ${{\mathbf{H}}_{be}} \sim \mathcal{C}\mathcal{N}\left( {0,M{{\left( {{\Xi _{be}}} \right)}^2}/2} \right)$. In this case, $|{{\mathbf{H}}_{be}}{|^2}$ follows chi-square distribution of 2 degrees of freedom and the negative exponential distribution with the given CDF in (\ref{cdf eve channel squ}). The proof is completed.
\end{proof}
\end{lemma}

\section{Secrecy Outage Probability Analysis}\label{SectionIV}
In this section, the SOP expressions for users in MF-RIS assisted NOMA networks are derived based on the channel statistical properties provided in Section \ref{SectionIII}. To be specific, the users's secrecy outage behaviours are evaluated under external and internal wiretapping scenarios with the practical consideration of imperfect SIC process as well as thermal noise.
\subsection{External Wiretapping Scenario}
\subsubsection{SOP Expressions of $u_r$ and $u_t$}
In the external wiretapping scenario, the E-Eve is not registered in the MF-RIS service cell and often remains silent, meaning that the BS cannot obtain its CSI and the channel diversity gain brought by the MF-RIS is also unavailable at the E-Eve. After the E-Eve obtains the confidential signal, SIC technique is employed to decode users' private message step by step. Hence, we define the secrecy capacity of $u_\varphi $ as $C_{ee,\varphi} ^\varsigma  = {\left[ {\log \left( {1 + \gamma _{\varphi ,\varphi }^\varsigma } \right) - \log \left( {1 + \gamma _{ee,\varphi }^\varsigma } \right)} \right]^ + }$, where $\varphi  \in \left\{ {r,t} \right\}$ and $\varsigma  \in \left\{ {{\text{ipSIC, pSIC}}} \right\}$. Note that SIC is not adopted when $u_r$ and E-Eve decode $u_t$'s information. Under this circumstance, the secrecy outage behaviours occur when ${C_\varphi ^\varsigma  < {R_\varphi }}$, and the SOP expressions are given in the following theorems.
\begin{theorem} \label{theorem1}
In the external wiretapping scenario, the SOP expressions for $u_r$ with imperfect SIC in MF-RIS-assisted NOMA networks is given in (\ref{sop ee r ipsic}) at the top of next page, where ${\mu _{rr1}} = {\rho _e}{a_r}{\chi ^2}{\beta _r}d_{b}^{ - \alpha }$, ${\mu _{rr2}} = \varpi {\rho _e}$, ${\mu _{rr3}} = {\beta _r}\chi \sigma _s^2M\left( {\left( {M{\kappa _{r}} + 1} \right)/\left( {{\kappa _{r}} + 1} \right)} \right)$, ${\varepsilon _{er1}} = {P_b}{a_r}{\chi ^2}{\beta _r}d_{e}^{ - \alpha }d_{b}^{ - \alpha }$, ${\varepsilon _{er2}} = \varpi {P_b}{\Omega_{e,r}^{ip}}$, ${\varepsilon _{er3}} = {\beta _r}\sigma _s^2\chi d_{e}^{ - \alpha }M\left( {\left( {M{\kappa _{e}} + 1} \right)/\left( {\left( {{\kappa _{e}} + 1} \right)\sigma _e^2} \right)} \right) + 1$, $q\left( {{z_w}} \right) = \left( {{z_w} + 1} \right){R_d}/2$, ${z_w} = \cos \left( {\left( {2w - 1} \right)\pi /\left( {2W} \right)} \right)$ with w = 1,2,3, $\cdots$ ,W, ${\Lambda _{r,ip}}\left( y \right) = {2^{{R_r}}}\left( {1 + {\varepsilon _{er1}}y/\left( {{\varepsilon _{er2}} + {\varepsilon _{er3}}} \right)} \right) - 1$ and we set ${\rho _e} = {P_b}/{n_e}$ indicating the SNR at the E-Eve. Moreover, ${G_x} = {\left( {X!} \right)^2}/\left( {{\tau _x}{{\left( {{{L'}_X}\left( {{\tau _x}} \right)} \right)}^2}} \right)$ and ${{\tau _x}}$ represent the weight of Gauss-Laguerre quadrature formula and the x-th zero point of Laguerre polynomial ${L_n}\left( x \right) = \sum\nolimits_{m = 0}^n {{{\left( { - 1} \right)}^m}} \left( {n!{x^m}} \right)/\left( {\left( {n - m} \right)!{{\left( {m!} \right)}^2}} \right)$ with x = 1,2,3, $\cdots$ ,X, respectively. X indicates a complexity accuracy tradeoff parameter , which is set to 300 to achieve a reliable approximation and the equal sign in (\ref{sop ee r ipsic}) can be established when X approaches infinity.
\begin{proof}
Please refer to Appendix A.
\end{proof}
\begin{figure*}[!t]
\begin{small}
\begin{align}\label{sop ee r ipsic}
P_{ee,r}^{ipSIC}\left( {{R_r}} \right) = \frac{\pi }{{W{R_d}}}\sum\limits_{s = 0}^S {\sum\limits_{d = 0}^D {\sum\limits_{w = 1}^W {\frac{{{G_s}{G_d}q\left( {{z_w}} \right)}}{{\Gamma \left( {{p_r}} \right)}}\sqrt {1 - z_w^2} } } } \gamma \left( {{k_r},\frac{1}{{{l_r}}}\sqrt {\left( {\left( {{\mu _{rr2}}{\Omega_{r}^{ip}}{\tau _s} + \sigma _n^2} \right){{\left( {q\left( {{z_w}} \right)} \right)}^\alpha } + {\mu _{rr3}}} \right){\Lambda _{r,ip}}\left( {M{\tau _d}} \right)/{\mu _{rr1}}} } \right).
\end{align}
\end{small}
\hrulefill \vspace*{0pt}
\end{figure*}
\end{theorem}

\begin{corollary} \label{corollary1}
Conditioned on $\varpi  = 0$, the SOP expression for $u_r$ with perfect SIC in MF-RIS-assisted NOMA networks is given by
\begin{small}
\begin{align}\label{sop ee r psic}
P_r^{psic}\left( {{R_r}} \right) =& \sum\limits_{d = 0}^D {\sum\limits_{w = 1}^W {\frac{{{G_d}q\left( {{z_w}} \right)\sqrt {1 - z_w^2} \pi }}{{W{R_d}\Gamma \left( {{p_r}} \right)}}} } \notag \\ &\times \gamma \left( {{k_r},\sqrt {\frac{{{\mu _{rr3}} + \sigma _n^2{{\left( {q\left( {{z_w}} \right)} \right)}^\alpha }}}{{l_r^2{\mu _{rr1}}{{\left( {\Lambda _r^p\left( {M{\tau _d}} \right)} \right)}^{ - 1}}}}} } \right),
\end{align}
\end{small}where ${\Lambda _{r,p}}\left( y \right) = {2^{{R_r}}}\left( {1 + {\varepsilon _{er1}}y/{\varepsilon _{er3}}} \right) - 1$.
\end{corollary}

Since the $u_t$ is regarded as the weak user in the external wiretapping scenario, more power will be allocated to it by the BS to guarantee its first decoding order without the SIC process.

\begin{theorem} \label{theorem2}
In the external wiretapping scenario, the SOP expression for $u_t$ in MF-RIS-assisted NOMA networks is denoted as
\begin{small}
\begin{align}\label{sop ee t}
{P_t}\left( {{R_t}} \right) =& \frac{\pi }{{W{R_d}}}\sum\limits_{d = 0}^D {\sum\limits_{w = 1}^W {\frac{{{G_d}q\left( {{z_w}} \right)\sqrt {1 - z_w^2} }}{{\Gamma \left( {{p_t}} \right)}}} }\notag \\ &\times \gamma \left( {{k_t},\sqrt {\frac{{\left( {{\mu _{tt3}} + \sigma _n^2{{\left( {q\left( {{z_w}} \right)} \right)}^\alpha }} \right){\Lambda _t}\left( {M{\tau _d}} \right)}}{{l_t^2\left( {{\mu _{tt1}} - {\mu _{tt2}}{\Lambda _t}\left( {M{\tau _d}} \right)} \right)}}} } \right),
\end{align}
\end{small}where ${\mu _{tt1}} = {P_b}{a_t}{\chi ^2}{\beta _t}d_{b}^{ - \alpha }$, ${\mu _{tt2}} = {P_b}{a_r}{\chi ^2}{\beta _t}d_{b}^{ - \alpha }$, ${\mu _{tt3}} = {\beta _t}\chi \sigma _s^2M\left( {M{\kappa _{t}} + 1} \right)/\left( {{\kappa _{t}} + 1} \right)$, ${\varepsilon _{et1}} = {\rho _e}{a_t}{\chi ^2}{\beta _r}d_{b}^{ - \alpha }d_{e}^{ - \alpha }$, ${\varepsilon _{et2}} = {\rho _e}{a_r}{\chi ^2}{\beta _r}d_{b}^{ - \alpha }d_{e}^{ - \alpha }$, ${\varepsilon _{et3}} = \left( {{\beta _r}\sigma _s^2\chi d_e^{ - \alpha }M\left( {M{\kappa _e} + 1} \right)/\left( {\sigma _e^2\left( {{\kappa _e} + 1} \right)} \right)} \right) + 1$, and ${\Lambda _t}\left( y \right) = {2^{{R_t}}}\left( {1 + {\varepsilon _{et1}}y/\left( {{\varepsilon _{et2}}y + {\varepsilon _{et3}}} \right)} \right) - 1$.
\end{theorem}

%

\subsubsection{Secrecy Diversity Order for $u_r$ and $u_t$}
In order to accurately assess the anti-eavesdropping fading capability of the MF-RIS-assisted NOMA networks, we further derive asymptotic SOP expressions of $u_t$ and $u_r$ at the high SNR region. Based on this, we obtain the secrecy diversity order that characterizes the robustness of users in the external eavesdropping scenario.

\begin{corollary} \label{corollary2}
When the transmitting SNR approaches infinity, the asymptotic SOP expression of $u_r$ with imperfect SIC in MF-RIS-assisted NOMA networks is given by
\begin{small}
\begin{align}\label{asy sop ee r ipsic}
P_{ee,r,\infty }^{ipSIC}\left( {{R_r}} \right) =& \sum\limits_{s = 0}^S {\sum\limits_{d = 0}^D {\sum\limits_{w = 1}^W {\frac{{\pi {G_s}{G_d}q\left( {{z_w}} \right)}}{{W{R_d}\Gamma \left( {{p_r}} \right)}}\sqrt {1 - z_w^2} } } }\notag \\ &\times \gamma \left( {{k_r},\frac{1}{{\hat \mu _{rr1}^2{l_r}}}\sqrt {\frac{{\varpi {\Omega_{r}^{ip}}{\tau _s}\Lambda _{r,ip}\left( {M{\tau _d}} \right)}}{{{{\left( {q\left( {{z_w}} \right)} \right)}^{ - \alpha }}}}} } \right),
\end{align}
\end{small}where ${{\hat \mu }_{rr1}} = {a_r}{\chi ^2}{\beta _r}d_{b}^{ - \alpha }$.
\begin{proof}
When the transmitting SNR approaches infinity, the SINR for $u_r$ to decode its own information with imperfect SIC can be recast as $\gamma _{r,r,\infty }^{ipSIC} = \frac{{{a_r}{\beta _r}{{\left| {{\mathbf{H}}_{r}^H{{\mathbf{\Phi }}_r}{{\mathbf{H}}_{b}}} \right|}^2}}}{{\varpi {{\left| {{\Omega_{r}^{ip}}} \right|}^2}}}$. Upon replacing $\gamma _{r,r}^{ipSIC}$ with $\gamma _{r,r,\infty }^{ipSIC}$ and referring the remaining steps in Appendix A, (\ref{asy sop ee r ipsic}) can be derived, and the proof is completed.
\end{proof}
\end{corollary}

\begin{corollary} \label{corollary3}
When the transmitting SNR approaches infinity, the asymptotic SOP expression of $u_r$ with perfect SIC in MF-RIS-assisted NOMA networks is given by (\ref{asy sop ee r psic}) shown at the top of next page, where $\Xi _{r,\infty }^{pSIC} = 16\left( {1 + {\kappa _{b}}} \right)\left( {1 + {\kappa _{r}}} \right)/\left( {3{e^{{\kappa _{b}} + {\kappa _{r}}}}} \right)$. The integral representation of hypergeometric functions is denoted as $_2{F_1}\left( {\alpha ,\beta ;\gamma ;z} \right) = \frac{1}{{{\rm B}\left( {\beta ,\gamma  - \beta } \right)}}\int_0^1 {{t^{\beta  - 1}}{{\left( {1 - t} \right)}^{\gamma  - \beta  - 1}}{{\left( {1 - tz} \right)}^{ - \alpha }}} dt$ \emph{\cite[Eq. (9.111)]{gradvstejn2000table}} and ${\rm B}\left( {x,y} \right) = 2{\int_0^1 {{t^{2x - 1}}\left( {1 - {t^2}} \right)} ^{y - 1}}dt$ represents the beta function \emph{\cite[Eq. (8.380.1)]{gradvstejn2000table}}.
\begin{figure*}[!t]
\begin{small}
\begin{align}\label{asy sop ee r psic}
P_{ee,r,\infty }^{psic}\left( {{R_r}} \right) = \frac{{\pi {{\left( {\Xi _{r,\infty }^{pSIC}} \right)}^M}}}{{\mu _{rr1}^MW{R_d}\left( {2M} \right)!}}{\left( {_2{F_1}\left( {2,\frac{1}{2};\frac{5}{2};1} \right)} \right)^M}\sum\limits_{d = 0}^D {\sum\limits_{w = 1}^W {{G_d}} } q\left( {{z_w}} \right)\sqrt {1 - z_w^2} {\left( {\left( {{\mu _{rr3}} + \sigma _n^2d_{r}^\alpha } \right)\Lambda _r^p\left( {M{\tau _d}} \right)} \right)^M}.
\end{align}
\end{small}
\hrulefill \vspace*{0pt}
\end{figure*}
\begin{proof}
Please refer to Appendix B.
\end{proof}
\end{corollary}

As for $u_t$, no SIC process is harnessed since it can decode its own signal directly. In this case, the asymptotic SOP expression of $u_t$ is given in the following corollary.
\begin{corollary} \label{corollary4}
For the case ${a_t} > {\Lambda _t}\left( {M{\tau _d}} \right){a_r}$ and ${a_t}  \leqslant  {\Lambda _t}\left( {M{\tau _d}} \right){a_r}$, the asymptotic SOP expressions of $u_t$ in MF-RIS-assisted NOMA networks are respectively given by
\begin{small}
\begin{align}\label{asy sop ee t 1}
{P_{ee,t,\infty }}\left( {{R_t}} \right) =& \frac{{\pi {{\left( {{\Xi _{t,\infty }^{pSIC}}} \right)}^M}}}{{W\left( {2M} \right)!{R_d}}}{\left( {_2{F_1}\left( {2,\frac{1}{2};\frac{5}{2};1} \right)} \right)^M} \notag \\ &\times \sum\limits_{d = 0}^D {\sum\limits_{w = 1}^W {{G_d}q\left( {{z_w}} \right)\sqrt {1 - z_w^2} } }  \notag \\ &\times {\left( {\frac{{\left( {{\mu _{tt3}} + \sigma _n^2{{\left( {q\left( {{z_w}} \right)} \right)}^\alpha }} \right){\Lambda _t}\left( {M{\tau _d}} \right)}}{{{\mu _{tt1}} - {\mu _{tt2}}{\Lambda _t}\left( {M{\tau _d}} \right)}}} \right)^M},
\end{align}
\end{small}
and
\begin{small}
\begin{align}\label{asy sop ee t 2}
{P_{ee,t,\infty }}\left( {{R_t}} \right) = 1,
\end{align}
\end{small}where ${\Xi _{t,\infty }^{pSIC}} = 16\left( {1 + {\kappa _{b}}} \right)\left( {1 + {\kappa _{t}}} \right)/\left( {3{e^{{\kappa _{b}} + {\kappa _{t}}}}} \right)$.
\end{corollary}

To gain more insights, the secrecy diversity order is adopted to evaluate the anti-eavesdropping fading characteristic of both users, which is defined as follows:
\begin{small}
\begin{align}\label{div order def}
D_{ee,\varphi }^\varsigma  =  - \mathop {\lim }\limits_{{P_b} \to \infty } \frac{{\log \left( {P_{ee,\varphi ,\infty }^\varsigma } \right)}}{{\log \left( {{P_b}} \right)}},
\end{align}
\end{small}where $\varsigma  \in \left\{ {{\text{ipSIC, pSIC}}} \right\}$ and $\varphi  \in \left\{ {r,t} \right\}$. The ${P_{ee,\varphi ,\infty }^\varsigma }$ can be obtained based on (\ref{asy sop ee r ipsic}), (\ref{asy sop ee r psic}), (\ref{asy sop ee t 1}), and (\ref{asy sop ee t 2}), respectively.

\begin{remark} \label{remark1}
By substituting (\ref{asy sop ee r psic}) and (\ref{asy sop ee t 1}) into (\ref{div order def}), the secrecy diversity orders of $u_r$ with perfect SIC and $u_t$ in the external wiretapping scenario are both equal to M, 
indicating that introducing more reconfigurable components in MF-RIS is beneficial for enhancing the anti-eavesdropping fading performance of the MF-RIS-assisted NOMA networks. Similarly, by substituting (\ref{asy sop ee r ipsic}) into (\ref{div order def}), we can obtain the secrecy diversity order of $u_r$ with imperfect SIC, which equals zero.  Although MF-RIS can support secure transmission in NOMA networks, the secrecy performance is still severely constrained by imperfect SIC in high SNR regions, leading to a noticeable error floor. The residual interference caused by ipSIC is positively correlated with the power level. Moreover, when the power budget is sufficient, it can be observed that the residual interference has a greater impact on network secrecy than the thermal noise generated by active devices. This phenomenon is also exhibited in Fig. \ref{sys_tp_thermal_ipsic}.
\end{remark}

\subsection{Internal Wiretapping Scenario}
\subsubsection{The SOP Expressions of $u_t$}
In the internal wiretapping scenario, the $u_r$ is considered as an I-Eve, and it may attempt to eavesdrop on the signal of legitimate $u_t$. It is worth noting that the I-Eve is a registered user in the MF-RIS service cell whose accurate CSI is known at the BS. Therefore, the channel gain of the I-Eve can benefit from the full channel diverity gains from the passive beamforming of MF-RIS, leading to enhanced wiretapping performance. This also makes the internal eavesdropping scenario more challenging than the external one. The secrecy capacity of $u_t$ in internal wiretapping scenario is denoted as $C_{ie,t}^\varsigma  = {\left[ {\log \left( {1 + \gamma _{t,t}^\varsigma } \right) - \log \left( {1 + \gamma _{ie,t}^\varsigma } \right)} \right]^ + }$, where $\varsigma  \in \left\{ {{\text{ipSIC, pSIC}}} \right\}$ and the condition for the secrecy outage event to occur is $C_{ie,t}^\varsigma < R_t$. Hence, the SOP expressions are given in the following theorems.

\begin{theorem} \label{theorem3}
In the internal wiretapping scenario, the SOP expression for $u_t$ with imperfect SIC in MF-RIS-assisted NOMA networks is given in (\ref{sop ie t ipsic}) shown at the top of next page, where ${\mu _{rt1}} = {P_b}{a_t}{\chi ^2}{\beta _t}d_{b}^{ - \alpha }$, ${\mu _{rt2}} = \varpi {P_b}{\Omega_{t}^{ip}}$, ${\mu _{rt3}} = {\beta _t}\chi \sigma _s^2M\left( {M{\kappa _{t}} + 1} \right)/\left( {{\kappa _{t}} + 1} \right)$, ${\varepsilon _{rt1}} = {P_b}{a_t}{\chi ^2}{\beta _r}d_{b}^{ - \alpha }$, ${\varepsilon _{rt2}} = \varpi {P_b}{\Omega_{r,t}^{ip}}$, ${\varepsilon _{rt3}} = {\beta _r}\chi \sigma _s^2M\left( {M{\kappa _{r}} + 1} \right)/\left( {{\kappa _{r}} + 1} \right)$, $h\left( x \right) = \left( {x + 1} \right)R_d^\alpha /2$, ${y_n} = \cos \left( {\left( {2n - 1} \right)\pi /\left( {2N} \right)} \right)$, ${z_i} = \cos \left( {\left( {2i - 1} \right)\pi /\left( {2I} \right)} \right)$ with $n = 1,2,3, \cdots ,N$, $i = 1,2,3, \cdots ,I$, $\Psi \left( z \right) = {2^{{R_t}}}\left( {1 + {\varepsilon _{rt1}}{\Omega _{br}}/\left( {\left( {{\varepsilon _{rt2}} + \sigma _n^2} \right)z + {\varepsilon _{rt3}}} \right)} \right) - 1$, and $\Xi _{ie,t}^{ipSIC} = {\pi ^2}R_d^{2\alpha  - 4}/\left( {IN{\alpha ^2}\Gamma \left( {{k_t}} \right)} \right)$. Note that the SOP expression for $u_t$ with perfect SIC in MF-RIS-assisted NOMA networks can be attained by setting $\varpi = 0$ in (\ref{sop ie t ipsic}).
\begin{figure*}[!t]
\begin{small}
\begin{align}\label{sop ie t ipsic}
P_{ie,t}^{ipSIC}\left( {{R_t}} \right) = \Xi _{ie,t}^{ipSIC}\sum\limits_{i = 1}^I {\sum\limits_{n = 1}^N {{{\left( {h\left( {{y_n}} \right)h\left( {{z_i}} \right)} \right)}^{\frac{2}{\alpha } - 1}}\sqrt {\left( {1 - y_n^2} \right)\left( {1 - z_i^2} \right)} } } \gamma \left( {{k_t},\frac{{\sqrt {\Psi \left( {h\left( {{z_i}} \right)} \right)\left( {\left( {{\mu _{rt2}} + \sigma _n^2} \right)h\left( {{y_n}} \right) + {\mu _{rt3}}} \right)} }}{{{l_t}\mu _{{\mu _{rt1}}}^2}}} \right).
\end{align}
\end{small}
\hrulefill \vspace*{0pt}
\end{figure*}
\begin{proof}
Please refer to Appendix C.
\end{proof}
\end{theorem}

\subsubsection{Secrecy Diversity Order for $u_t$}
In order to reap further insights, the asymptotic expressions of SOP for $u_t$ with imperfect SIC under high SNR region is derived, which is shown in the following corollary.
\begin{corollary} \label{corollary5}
When the transmitting SNR approaches infinity, the asymptotic expression of SOP for $u_t$ with imperfect SIC in MF-RIS-assisted NOMA networks is given by
\begin{small}
\begin{align}\label{asy sop ie t ipsic}
P_{ie,t,\infty }^{ipSIC}\left( {{R_t}} \right) =& \Xi _{ie,t}^{ipSIC}\sum\limits_{i = 1}^I {\sum\limits_{n = 1}^N {{{\left( {h\left( {{y_n}} \right)h\left( {{z_i}} \right)} \right)}^{\frac{2}{\alpha } - 1}}} }\notag \\ &\times {\left( {\left( {1 - y_n^2} \right)\left( {1 - z_i^2} \right)} \right)^{\frac{1}{2}}} \notag \\ &\times \gamma \left( {{k_t},\frac{1}{{{l_t}\hat \mu _{rt1}^2}}\sqrt {\Upsilon \left( {h\left( {{z_i}} \right)} \right){\hat \mu _{rt2}}h\left( {{y_n}} \right)} } \right),
\end{align}
\end{small}where ${{\hat \mu }_{rt1}} = {a_t}{\chi ^2}{\beta _t}d_{b}^{ - \alpha }$, ${{\hat \mu }_{rt2}} = \varpi {\Omega _{t}^{ip}}$, ${{\hat \varepsilon }_{rt1}} = {a_t}{\chi ^2}{\beta _r}d_{b}^{ - \alpha }$, ${{\hat \varepsilon }_{rt2}} = \varpi {\Omega _{r}^{ip}}$, and $\Upsilon \left( z \right) = {2^{{R_t}}}\left( {1 + {\hat \varepsilon _{rt1}}{\Omega _{br}}/\left( {{\hat \varepsilon _{rt2}}z} \right)} \right) - 1$.
\end{corollary}

\begin{corollary} \label{corollary6}
When the transmitting SNR approaches infinity, the asymptotic expression of SOP for $u_t$ with perfect SIC in MF-RIS-assisted NOMA networks is given by
\begin{small}
\begin{align}\label{asy sop ie t psic}
P_{ie,t,\infty }^{pSIC}\left( {{R_t}} \right) =& \frac{{{{\left( {\Xi _{t,\infty }^{pSIC}} \right)}^M}{\pi ^2}R_d^{2\alpha  - 4}}}{{NI{\alpha ^2}\left( {2M} \right)!}}{\left( {_2{F_1}\left( {2,\frac{1}{2};\frac{5}{2};1} \right)} \right)^M}\notag \\ &\times \sum\limits_{i = 1}^I {\sum\limits_{n = 1}^N {h{{\left( {{z_i}} \right)}^{\frac{2}{\alpha } - 1}}{{\left( {\Delta \left( {h\left( {{z_i}} \right),h\left( {{y_n}} \right)} \right)} \right)}^M}} }  \notag \\  &\times h{\left( {{y_n}} \right)^{\frac{2}{\alpha } - 1}}\sqrt {\left( {1 - y_n^2} \right)\left( {1 - z_i^2} \right)},
\end{align}
\end{small}where $\Xi _{t,\infty }^{pSIC} = 16\left( {1 + {\kappa _{b}}} \right)\left( {1 + {\kappa _{t}}} \right)/\left( {3{e^{{\kappa _{b}} + {\kappa _{t}}}}} \right)$ and $\Delta \left( {z,y} \right) = \Psi \left( z \right)\left( {\sigma _n^2y + {\mu _{rt3}}} \right)/{\mu _{rt1}}$.
\end{corollary}

\begin{remark} \label{remark2}
By substituting (\ref{asy sop ie t ipsic}) and (\ref{asy sop ie t psic}) into (\ref{div order def}), the secrecy diversity order of $u_t$ with imperfect and perfect SIC in the internal wiretapping scenario are both equal to zero. The reason is that 
cascaded channels of $u_t$ and I-Eve can benefit from the beamforming gain provided by MF-RIS. Therefore, the secrecy diversity order of $u_t$ is no longer dependent on the number of reconfigurable elements. In addition, due to the residual interference caused by ipSIC being directly proportional to the power level, the continuous increase in power budget has minimal effect on improving the user's SOP. This phenomenon is illustrated in Fig. \ref{IE_sop_diff_power}, indicating that the internal wiretapping scenario poses a greater challenge for MF-RIS-secured NOMA networks.
\end{remark}

\subsection{Secrecy Throughput Analysis}
According to the aforementioned SOP performance, the delay-limited throughput of $u_\varphi $ in MF-RIS-secured NOMA networks is defined as
\begin{small}
\begin{align}\label{tp_define}
\mathcal{T}_{\zeta ,\varphi }^\varsigma \left( {{R_\varphi }} \right) = \left( {1 - P_{\zeta ,\varphi }^\varsigma \left( {{R_\varphi }} \right)} \right){R_\varphi },
\end{align}
\end{small}where $\zeta  \in \left\{ {ee,ie} \right\}$, $\varphi  \in \left\{ {t,r} \right\}$ and $\varsigma  \in \left\{ {{\text{ipSIC, pSIC}}} \right\}$ \cite{Yingjie2023RISPLS},\cite{Xinwei2023STARRIS}. ${P_{\zeta ,\varphi }^\varsigma \left( {{R_\varphi }} \right)}$ can be attained from (\ref{sop ee r ipsic}), (\ref{sop ee r psic}), (\ref{sop ee t}), and (\ref{sop ie t ipsic}).

\section{Numerical Results}\label{SectionV}

In this section, numerical results are provided to verify the accuracy of the theoretical expressions derived in Section \ref{SectionIV}. Distinct secrecy insights of MF-RIS-assisted NOMA networks are also discussed in detail. Without otherwise stated, we set $\alpha  = 2.2$, $\chi  =  - 30$ dB \cite{Yue2024ASTRSNOMA}. The distance parameters are set to be ${R_d} = 20$ m, ${d_{b}} = 200$ m, and ${d_{e}} = 15$ m.  The amplification factors at MF-RIS are $\beta_r = \beta_t = 10 $ dB. Supposing that residual interference is approximately equivalent to 10\% of the power of the decoded signal, the imperfect SIC parameter in external and internal wiretapping scenarios can be set as $\Omega _r^{ip} \approx  - 126.13$ dB, $\Omega _{e,r}^{ip} \approx  - 130.26$ dB, $\Omega _t^{ip} \approx  - 125.04$ dB and $\Omega _{r,t}^{ip} \approx  - 129.17$ dB, respectively \cite{Yue2024ASTRSNOMA}. The power setup of AWGN and thermal noise are given as $\sigma _n^2 = \sigma _e^2 =  - 90$ dBm and $\sigma _s^2 = - 80$ dBm, respectively \cite{Ailing2023MFRIS},\cite{Ailing2023NextGenMFRIS}. In addition, the target secrecy rates for users $R_r = 0.1$ and $R_t = 0.05$ bit per channel use (BPCU) \cite{Yingjie2023ARISNOMA}. The power consumed by phase shifters and direct current biasing circuits are ${P_{ps}} = - 10$ and ${P_{dc}} = - 5$ dBm \cite{2022CunhuaPanARIS}, respectively. To highlight the secrecy performance of the proposed network, the active RIS and STAR-RIS are both considered as benchmarks, which are defined as follows.

\begin{enumerate}
  \item \textbf{Active RIS}: Active RIS allows $M/2$ reconfigurable elements to perform only reflection functions, while the other $M/2$ elements exclusively perform refraction functions, and all the elements can independently implement phase modulation. Additionally, active RIS has signal amplification capabilities and each reconfigurable element is equipped with a set of reflection or refraction power amplifiers. We suppose that the elements have identical amplification factors, i.e., $\beta _{t,m} = \beta _{r,m},\beta _{\varphi,1} = \beta _{\varphi,2} =  \cdots  = \beta _{\varphi,{M/2}}$ and $\varphi  \in \left\{ {t,r} \right\}$.
  \item \textbf{STAR-RIS}: STAR-RIS can also perform dual-sided signal transmission, based on an ES strategy, simultaneously radiating the modulated signals to both the reflection and refraction regions. Unlike MF-RIS, STAR-RIS lacks signal amplification capabilities and does not include any active devices related to power amplification. In other words, STAR-RIS is almost passive. To be fair, the ES coefficient for STAR-RIS is set to be the same as that for MF-RIS, i.e., ${e_r}$ = 0.8 and ${e_t}$ = 0.2.
\end{enumerate}

\subsection{External Wiretapping Scenario}

\begin{figure}[t!]
    \begin{center}
        \includegraphics[width=3in,  height=2.4in]{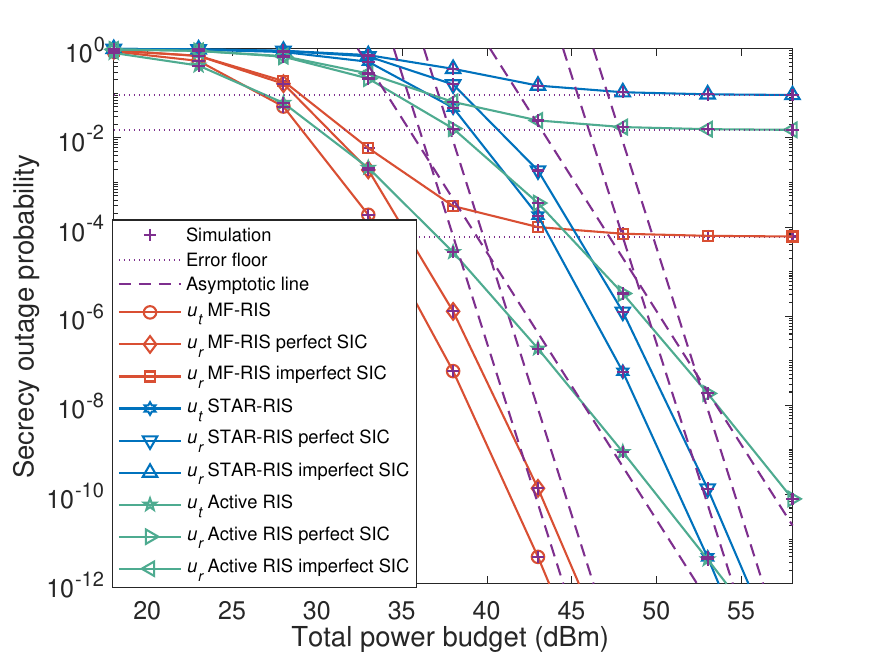}
        \caption{The SOP performance versus the total power budget in the external wiretapping scenario, where \emph{M} = 12, ${e_r}$ = 0.8, ${e_t}$ = 0.2, ${a_r}$ = 0.25 and ${a_t}$ = 0.75.}
        \label{sop_diff_power}
    \end{center}
\end{figure}

Fig. \ref{sop_diff_power} plots the SOP performance versus the system power budget in the external eavesdropping scenario. 
It is clear to see that the obtained analytical expressions are consistent with simulation results during the entire range of power budget, which validates the accuracy of analytical methods applied. Within the high SNR region, trends of the theoretical lines gradually converge with the asymptotic ones based on (\ref{asy sop ee r ipsic}), (\ref{asy sop ee r psic}) and (\ref{asy sop ee t 1}) manifesting the validity of asymptotic theoretical expressions. Since the slopes of the asymptotic lines for MF-RIS and STAR-RIS are both greater than that of the active RIS, this phenomenon also corroborates the analysis in Remark \ref{remark1}. From Fig. \ref{sop_diff_power}, we can observe that under the same system power constraints, the SOP performance of the proposed MF-RIS scheme significantly outperforms the STAR-RIS and active RIS alternatives. This is primarily due to the inherent advantages of MF-RIS, which encompass full-space coverage and signal amplification capabilities, capable of mitigating the adverse effects of multiplicative fading while enabling 360-degree wireless transmission\footnote{Although MF-RIS offers more rich and flexible design freedom, under certain conditions―such as limited power budget or short distance between network nodes, MF-RIS consumes additional power but fails to fully leverage its advantages. In these scenarios, the low-complexity active RIS and the passive STAR-RIS can provide higher gains for secure communication in NOMA networks.}. Additionally, Fig. \ref{sop_diff_power} illustrates the impact of imperfect SIC on the physical layer secrecy for $u_r$. For active RIS, STAR-RIS and MF-RIS, the residual interference generated by imperfect SIC forces the SOP curves for $u_r$ to converge to different error floors. This indicates that in the high SNR region, residual interference poses a more severe threat to users' privacy than thermal noise. 

\begin{figure}[t!]
    \begin{center}
        \includegraphics[width=3in,  height=2.4in]{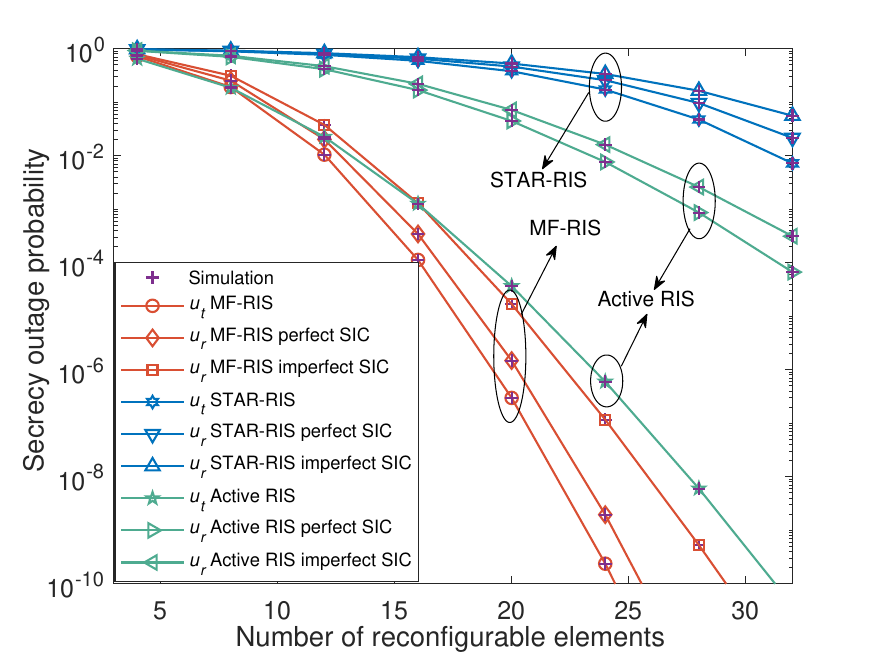}
        \caption{The SOP performance versus the number of reconfigurable elements in the external wiretapping scenario, where ${e_r}$ = 0.8, ${e_t}$ = 0.2, ${a_r}$ = 0.3 and ${a_t}$ = 0.7.}
        \label{sop_diff_m}
    \end{center}
\end{figure}

Fig. \ref{sop_diff_m} plots the SOP performance versus the number of reconfigurable elements in the external eavesdropping scenario. Similar to Fig. \ref{sop_diff_power}, the numerical results align closely with the theoretical expressions. On the one hand, it can be observed that increasing the number of elements is advantageous in reducing the SOP for users in both the reflection and refraction regions. On the other hand, owing to MF-RIS's capability for independent bidirectional signal manipulation and amplification, it can achieve equivalent secrecy performance with the fewest reconfigurable elements compared to active RIS and STAR-RIS. Furthermore, we can note that the presence of imperfect SIC diminishes the performance gains obtained from MF-RIS elements. Therefore, devising appropriate SIC receivers to mitigate the impact of residual interference is crucial for enhancing the physical layer secrecy of MF-RIS-assisted NOMA networks.

\begin{figure}[t!]
    \begin{center}
        \includegraphics[width=3in,  height=2.4in]{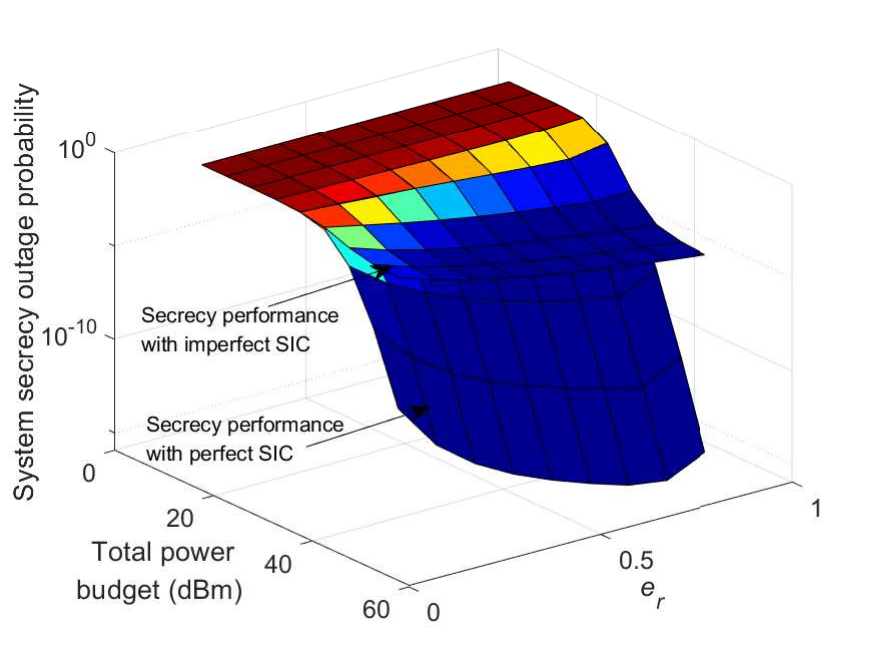}
        \caption{The SOP performance versus the total power budget and ES coefficients in the external wiretapping scenario, where \emph{M} = 12, ${a_r}$ = 0.25 and ${a_t}$ = 0.75.}
        \label{sop_diff_noise_3d}
    \end{center}
\end{figure}

Fig. \ref{sop_diff_noise_3d} plots the system SOP performance versus the total power budget and the ES parameter in the external eavesdropping scenario. As the system power budget increases, the performance of the MF-RIS-assisted NOMA networks under perfect SIC improves steadily in terms of SOP. However, in the imperfect SIC case, the system SOP gradually converging to an error floor since the residual interference escalates with higher power budget. 
Furthermore, it is evident that when MF-RIS allocates more power to $u_r$ (i.e., $e_{r} > 0.5$), the system secrecy outage behavior becomes better. 
However, when the $e_r$ exceeds a certain threshold, the system's SOP increases instead of decreasing. This is because that an excessively large $e_r$ leads to a reduction in the energy allocated to $u_t$, consequently diminishing the SINR  when $u_t$ decoding its own signal, ultimately compromising its secrecy capacity. This underscores the critical importance of selecting an appropriate ES strategy at the MF-RIS to safeguard NOMA networks.

\begin{figure}[t!]
    \begin{center}
        \includegraphics[width=3in,  height=2.4in]{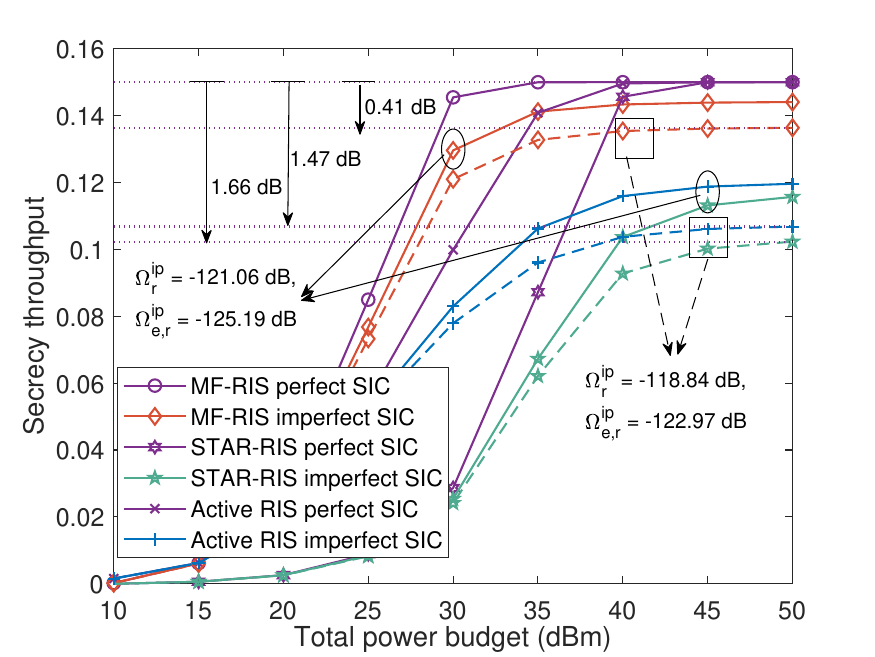}
        \caption{The secrecy throughput performance versus the total power budget in the external wiretapping scenario, where \emph{M} = 12, ${e_r}$ = 0.8, ${e_t}$ = 0.2, ${a_r}$ = 0.25 and ${a_t}$ = 0.75.}
        \label{sys_tp_diff_power}
    \end{center}
\end{figure}

Fig. \ref{sys_tp_diff_power} plots the system secrecy throughput versus the total power budget in the external eavesdropping scenario. From the figure, it can be observed that the delay-limited secrecy throughput of the MF-RIS is consistently higher than those of the benchmarks and eventually tends to stabilize with the increasing total power budget under perfect SIC case. This is because the SOP for users in the NOMA network has already become sufficiently small within high power budget. The residual interference caused by imperfect SIC has the least impact on the proposed MF-RIS scheme, with a throughput loss of approximately 0.41 dB. In contrast, active RIS and STAR-RIS are more significantly affected by imperfect SIC, with throughput losses of up to 1.47 dB and 1.66 dB, respectively. This indicates that the MF-RIS demonstrates excellent robustness compared to the other two RIS configuration schemes.

%

\begin{figure}[t!]
    \begin{center}
        \includegraphics[width=3in,  height=2.4in]{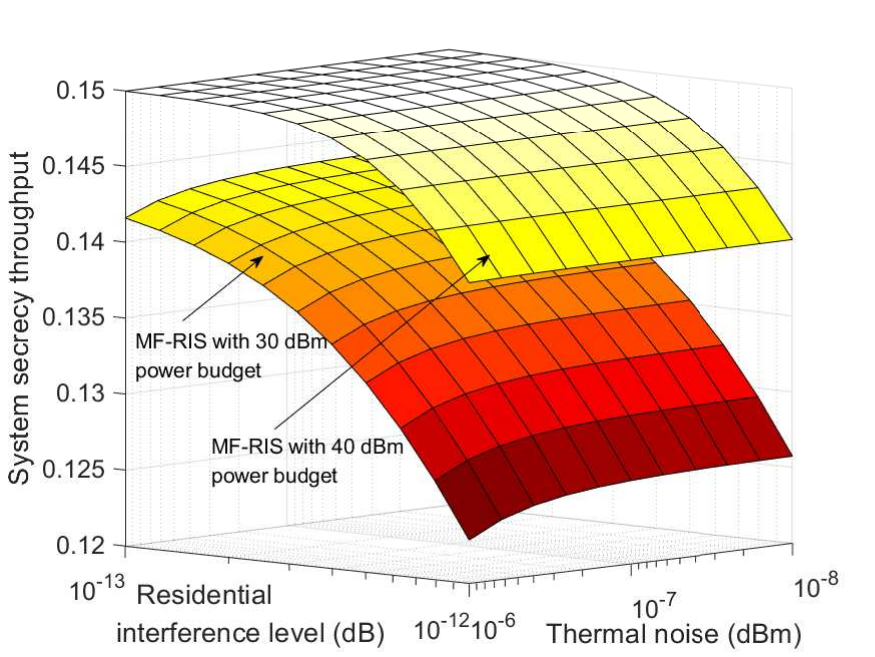}
        \caption{The secrecy throughput performance versus the residential interference and thermal noise in the external wiretapping scenario, where \emph{M} = 12, ${e_r}$ = 0.8, ${e_t}$ = 0.2, ${a_r}$ = 0.25 and ${a_t}$ = 0.75.}
        \label{sys_tp_thermal_ipsic}
    \end{center}
\end{figure}

Fig. \ref{sys_tp_thermal_ipsic} plots the system secrecy throughput versus the residential interference and thermal noise in the external eavesdropping scenario. This figure illustrates that increasing the total power budget can indeed enhance the overall secrecy throughput. When the total power budget is relatively abundant (40 dBm), the secrecy throughput of the MF-RIS-assisted NOMA networks is primarily influenced by the residual interference as the degree of imperfect SIC is directly proportional to the transmitting power of the BS. On the contrary, when the total power budget decreases (30 dBm), the secrecy throughput is jointly determined by the residual interference as well as the thermal noise. The phenomenon illustrated in Fig. \ref{sys_tp_thermal_ipsic} also answers the first and second questions that we raised in the introduction part, indicating that further optimization of SIC performance is crucial for maximizing secrecy in high-power regimes. These insights suggest that the design of MF-RIS-secured NOMA networks should carefully balance power allocation and SIC performance to enhance secrecy.

\subsection{Internal Wiretapping Scenario}

\begin{figure}[t!]
    \begin{center}
        \includegraphics[width=3in,  height=2.4in]{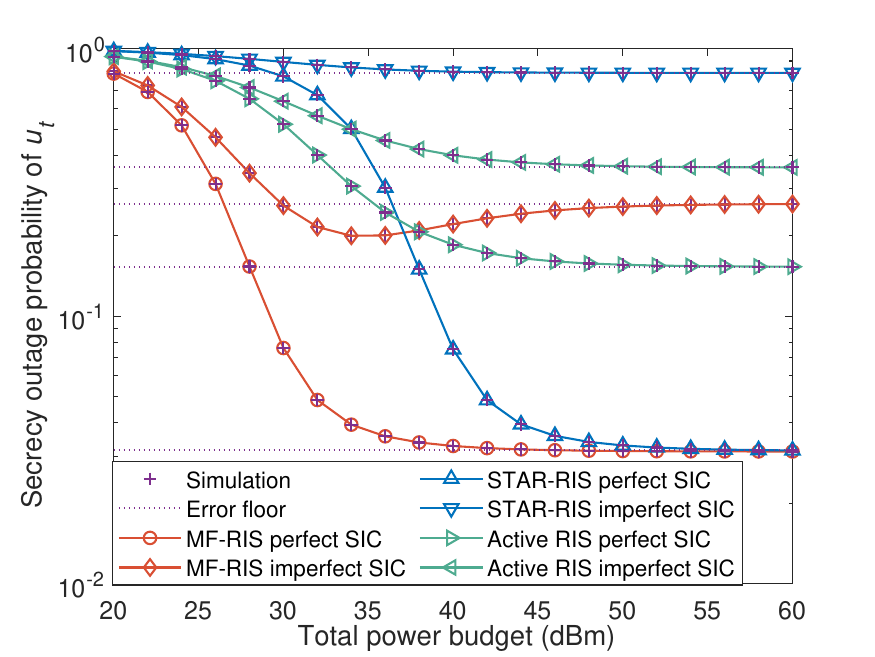}
        \caption{The SOP performance versus the total power budget in the internal wiretapping scenario, where \emph{M} = 12, ${e_r}$ = 0.2, ${e_t}$ = 0.8, ${a_r}$ = 0.9 and ${a_t}$ = 0.1.}
        \label{IE_sop_diff_power}
    \end{center}
\end{figure}

Fig. \ref{IE_sop_diff_power} plots the SOP performance versus the total power budget in the internal eavesdropping scenario. From the figure, we can observe that the secrecy outage behavior for $u_t$ in MF-RIS-assisted NOMA networks is superior to the other two benchmarks in both imperfect and perfect SIC cases. Furthermore, unlike the external eavesdropping scenario, the SOP for $u_t$ eventually converges instead of continuing to decrease under perfect SIC conditions, and this phenomenon also corroborates the analyses in Remark \ref{remark2}. This is because the I-Eve's received SNR is proportional to the transmit power of BS. Therefore, as the total power budget increases, the wiretapping capability of the I-Eve also improves. Moreover, it can be seen that under the imperfect SIC condition, the SOP for $u_t$ initially decreases and then increases. This can be explained by the fact that in the high SNR region, the residual interference introduced by imperfect SIC is higher. At this point, the impact of residual interference on legitimate $u_t$ is greater compared to its impact on I-Eve, leading to fluctuations in the SOP curve.

\begin{figure}[t!]
    \begin{center}
        \includegraphics[width=3in,  height=2.4in]{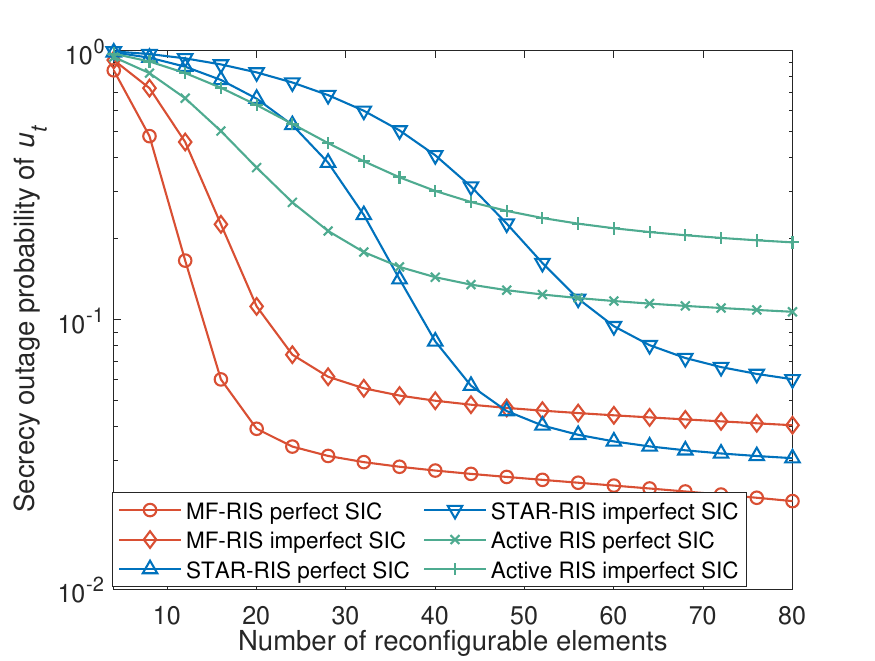}
        \caption{The SOP performance versus the number of reconfigurable elements in the internal wiretapping scenario, where ${e_r}$ = 0.2, ${e_t}$ = 0.8, ${a_r}$ = 0.9 and ${a_t}$ = 0.1.}
        \label{IE_sop_diff_m}
    \end{center}
\end{figure}

Fig. \ref{IE_sop_diff_m} plots the SOP performance versus the number of reconfigurable elements in the internal eavesdropping scenario. With the number of elements increases, the SOP of $u_t$ gradually decreases, and the rate of decrease slows down. This is because, with a greater number of reconfigurable elements, the quality of cascaded eavesdropping channels for the I-Eve strengthens. In other words, both $u_t$ and I-Eve can benefit from the beamforming gains from MF-RIS, which is evidently different from the external eavesdropping scenario illustrated in Fig. \ref{sop_diff_m}. Additionally, the figure indicates that MF-RIS achieves the same level of secrecy performance as active RIS and STAR-RIS with only a small number of elements. This is attributed to the high diversity order and dual-sided signal amplification capability of MF-RIS. The SOP performances shown in Fig. \ref{IE_sop_diff_power} and Fig. \ref{IE_sop_diff_m} are remarkably different from that in the external wiretapping scenario, which also answers the last question proposed in the introduction part.

\begin{figure}[t!]
    \begin{center}
        \includegraphics[width=3in,  height=2.4in]{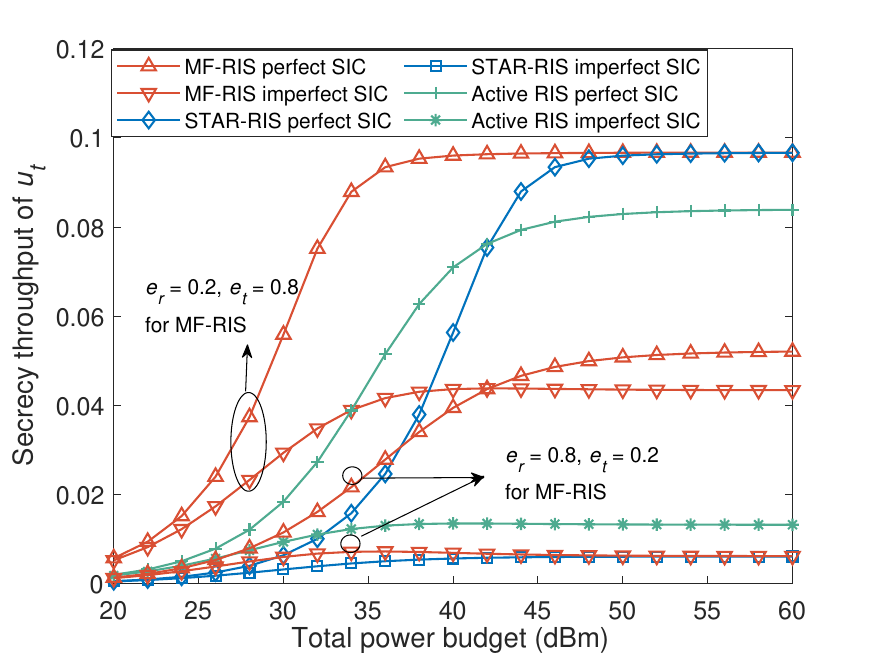}
        \caption{The secrecy throughput performance versus the total power budget in the internal wiretapping scenario, where ${a_r}$ = 0.9, ${a_t}$ = 0.1 and ${R_t}$ = 0.1 BPCU.}
        \label{IE_tp}
    \end{center}
\end{figure}

Fig. \ref{IE_tp} plots the secrecy throughput versus the total power budget in the internal eavesdropping scenario. It can be seen from the figure that
the secrecy throughput of $u_t$ in the MF-RIS-assisted NOMA networks surpasses that of the other two benchmarks, mirroring the user's secrecy throughput performance in the external eavesdropping scenario, as illustrated in Fig. \ref{sys_tp_diff_power}. It is noteworthy that the method to enhance the communication secrecy for the $u_t$ involves deferring the decoding order of confidential information until it is the last one. Furthermore, MF-RIS should split more energy to $u_t$, specifically with $e_r = 0.2$ and $e_t = 0.8$. This ES strategy aims to diminish the SINR when the I-Eve decodes the confidential information, thereby augmenting the secrecy capacity of $u_t$. In cases where such assurance is unattainable, i.e., with $e_r = 0.8$ and $e_t = 0.2$, as evidenced in the figure, the secrecy throughput of $u_t$ in the internal eavesdropping scenario experiences a notable degradation.

\section{Conclusion} \label{SectionVI}
In this paper, we provided both theoretical insights and practical guidelines for analyzing the physical layer secrecy of MF-RIS-assisted NOMA networks operating in ES mode under both external and internal eavesdropping scenarios. We derived the SOP and secrecy throughput expressions of randomly distributed users, and subsequently obtained the secrecy diversity orders in high SNR region, which offered a framework for guiding decisions on the number of MF-RIS elements, power budget design, and handling of residual interference. Theoretical analyses indicated that the secrecy diversity order of users is proportional to the number of reconfigurable elements only when the users are in the external eavesdropping scenario or ideal SIC case. Numerical results firstly demonstrated the superior secrecy performance of MF-RIS-assisted NOMA networks compared to existing RIS counterparts. Secondly, residual interference caused by imperfect SIC posed a greater threat to the MF-RIS-secured NOMA networks with incremental total power budget. Finally, by redistributing the power allocation at BS and the ES coefficients at MF-RIS, the secrecy rates of users can be guaranteed in the internal wiretapping scenario.
\section{Acknowledgement}
The authors would like to express their sincere gratitude to the Editor the anonymous Reviewers for their insightful comments and constructive suggestions, which have significantly improved the quality and clarity of this paper. Additionally, the work of Dr. Wanli Ni presented in this paper, especially the core concept of MF-RIS, was conducted during his doctoral studies at Beijing University of Posts and Telecommunications.

\appendices
\section*{Appendix~A: Proof of Lemma \ref{Lemma1}} \label{AppendixA}
\renewcommand{\theequation}{A.\arabic{equation}}
\setcounter{equation}{0}
In the external wiretapping scenario, the secrecy outage event of $u_t$ occurs when $C_r^{ipSIC} = {\left[ {\log \left( {1 + \gamma _{r,r}^{ipSIC}} \right) - \log \left( {1 + \gamma _{ee,r}^{ipSIC}} \right)} \right]^ + } < {R_r}$. As a consequence, the SOP expression of $u_t$ with imperfect SIC is denoted as
\begin{small}
\begin{align}\label{a1}
P_{ee,r}^{ipSIC} = \Pr \left( {{{\left( {\log \left( {\frac{{1 + \gamma _{r,r}^{ipSIC}}}{{1 + \gamma _{ee,r}^{ipSIC}}}} \right)} \right)}^ + } < {R_r}} \right),
\end{align}
\end{small}where ${\gamma _{r,r}^{ipSIC}}$ and ${\gamma _{ee,r}^{ipSIC}}$ are given in (\ref{SINR r decode r ipSIC}) and (\ref{SINR EE decode r ipsic}), respectively. Note that the thermal noise is ${\left| {{\mathbf{H}}_{r}^H{{\mathbf{\Phi }}_r}{{\mathbf{n}}_s}} \right|^2} = \chi d_{r}^{ - \alpha }\sigma _s^2{\left| {\sum\nolimits_{m = 1}^M {h_{r}^m} {e^{j\theta _r^m}}} \right|^2}$. To facilitate the calculation process, we define ${\Delta _m} = h_{r}^m{e^{j\theta _r^m}}$ and ${\Delta _m}$ is regarded as a novel complex Gaussian random variable which satisfies ${\Delta _m} \sim \mathcal{C}\mathcal{N}\left( {a,2{\sigma ^2}} \right)$. In this case, the expectation as well as variance operation can be shown as
\begin{small}
\begin{align}\label{a3}
\mathbb{E}\left( {{{\left| {{\Delta _m}} \right|}^2}} \right) = \mathbb{D}\left( {\left| {{\Delta _m}} \right|} \right) + {\left( {\mathbb{E}\left( {\left| {{\Delta _m}} \right|} \right)} \right)^2},
\end{align}
\end{small}
and
\begin{small}
\begin{align}\label{a4}
\mathbb{E}\left( {{{\left| {{\Delta _m}} \right|}^2}} \right) =& \left( {{M^2}{\kappa _{r}}/\left( {{\kappa _{r}} + 1} \right)} \right) + \left( {M/\left( {{\kappa _{r}} + 1} \right)} \right) \notag \\  =& M\left( {M{\kappa _{r}} + 1} \right)/\left( {{\kappa _{r}} + 1} \right),
\end{align}
\end{small}respectively. Hence, the thermal noise received at $u_r$ and E-Eve can be replaced with $M\left( {M{\kappa _{r}} + 1} \right)/\left( {{\kappa _{r}} + 1} \right)$ and $M\left( {M{\kappa _{e}} + 1} \right)/\left( {{\kappa _{e}} + 1}  \right)$. We further assume that $X = {\left| {{{{\mathbf{\hat H}}}_{br}}} \right|^2}$, $Y = {\left| {{\mathbf{h}}_{e}^H{{\mathbf{\Phi }}_r}{{\mathbf{h}}_{b}}} \right|^2}$, $Z = {d_{r}}$ and $V = {\left| {{h_{ipu}}} \right|^2}$. The (\ref{a1}) can be derived as
\begin{small}
\begin{align}\label{a5}
P_{ee,r}^{ipSIC} =& \int_0^\infty  {\int_0^\infty  {\int_0^{{R_d}} {{f_Z}\left( z \right)} {f_V}\left( v \right){f_Y}\left( y \right)dydvdz} }  \notag \\ &\times {F_X}\left( {\frac{{\Lambda _{r,ip}\left( y \right)}}{{{\mu _{rr1}}}}\left( {{\mu _{rr2}}vd_{r}^\alpha  + {\mu _{rr3}} + \sigma _n^2d_{r}^\alpha } \right)} \right).
\end{align}
\end{small}

In order to derive tractable and analytically convenient expressions for SOP and secrecy throughput, we employed Gaussian-Laguerre approximations and Gauss-Chebyshev quadrature to solve the complex integrals in (\ref{a5}). By using these well-established mathematical techniques, we reduce the complexity of the original integrals, making it feasible to extract key insights about the performance of MF-RIS-secured networks. Defining ${\Delta _1} = \int_0^{{R_d}} {{f_Z}\left( z \right)} {F_X}\left( {\Lambda _{r,ip}\left( y \right)\left( {{\mu _{rr2}}vd_{r}^\alpha  + {\mu _{rr3}}} \right.} \right.$ $ + \left. {\sigma _n^2d_{r}^\alpha } \right)\left. {/{\mu _{rr1}}} \right)$ and substituting (\ref{pdf drr drt}) into it, the ${\Delta _1}$ can be rewritten with the assistance of Gauss-Chebyshev quadrature \cite[Eq. (8.8.4)]{hildebrand1987introduction} as follows
\begin{small}
\begin{align}\label{a6}
{\Delta _1} =& \frac{\pi }{{\Gamma \left( {{k_r}} \right)W{R_d}}}\sum\limits_{w = 1}^W {q\left( {{z_w}} \right)} \sqrt {1 - z_w^2}\notag \\ &\times \gamma \left( {{k_r},\frac{{\sqrt {\left( {{\mu _{rr2}}v + \sigma _n^2} \right){{\left( {q\left( {{z_w}} \right)} \right)}^\alpha } + {\mu _{rr3}}} }}{{{l_r}\sqrt {{\mu _{rr1}}{{\left( {\Lambda _{r,ip}\left( y \right)} \right)}^{ - 1}}} }}} \right).
\end{align}
\end{small}Considering ${f_{{{\left| {{h_\varphi^{ip}}} \right|}^2}}}\left( v \right) = \left( {1/{\Omega_\varphi^{ip}}} \right){e^{ - y/{\Omega_\varphi^{ip}}}}$ \cite{Yingjie2023RISPLS}, the (\ref{a5}) can be rewritten as
\begin{small}
\begin{align}\label{a7}
P_{ee,r}^{ipSIC} = \frac{\pi }{{MW{R_d}}}\int_0^\infty  {{f_V}\left( v \right)dv} \int_0^\infty  {{e^{ - \frac{1}{M}y}}} {\Delta _1}.
\end{align}
\end{small}By harnessing the Gauss-Laguerre quadrature formula and some straightforward calculations \cite[Eq. (8.6.5)]{hildebrand1987introduction}, the integral in (\ref{a7}) can be calculated to obtain (\ref{sop ee r ipsic}), which completes the proof.

\section*{Appendix~B: Proof of Corollary \ref{corollary3} } \label{AppendixB}
\renewcommand{\theequation}{B.\arabic{equation}}
\setcounter{equation}{0}
To obtain the asymptotic SOP for $u_r$, it is pivotal to explore the statistical properties of ${{{{\mathbf{\hat H}}}_{b\varphi }}}$ when the transmit power approaches infinity. According to \cite[Eq. (6.621.3)]{gradvstejn2000table}, the PDF of ${\left| {h_{\varphi,m}h_{b,m}} \right|}$ can be given in (\ref{b1-1}) via utilizing Laplace transform as shown at the top of this page, where 
$m = 1,2, \cdots ,M$ and $\varphi  \in \left\{ {t,r} \right\}$.
\begin{figure*}[!t]
\begin{small}
\begin{align}\label{b1-1}
\mathcal{L}\left( {{f_{\left| {h_{\varphi,m}h_{b,m}} \right|}}\left( x \right)} \right)\left( s \right) =& \sum\limits_{a = 0}^\infty  {\sum\limits_{b = 0}^\infty  {\frac{{{{\left( {{\kappa _{b}}{\kappa _{\varphi }}} \right)}^a}{{\left( {\left( {1 + {\kappa _{b}}} \right)\left( {1 + {\kappa _{\varphi }}} \right)} \right)}^{\left( {a + 1} \right)}}}}{{{4^{b - a - 1}}{\pi ^{ - \frac{1}{2}}}{{\left( {\left( {a!} \right)\left( {b!} \right)} \right)}^2}{e^{{\kappa _{b}} + {\kappa _{\varphi }}}}}}} }\notag \\ &{ \times _2}{F_1}\left( {2\left( {a + 1} \right),\frac{1}{2} + a - b;\frac{5}{2} + a + b;\frac{{s - 2\sqrt {\left( {1 + {\kappa _{b}}} \right)\left( {1 + {\kappa _{\varphi }}} \right)} }}{{s + 2\sqrt {\left( {1 + {\kappa _{b}}} \right)\left( {1 + {\kappa _{\varphi }}} \right)} }}} \right)\Gamma \left( \begin{gathered}
  2\left( {b + 1} \right),2\left( {a + 1} \right) \hfill \\
  {\text{\quad\quad}}a + b + \frac{5}{2} \hfill \\
\end{gathered}  \right)
\end{align}
\end{small}
\hrulefill \vspace*{0pt}
\end{figure*}Under the condition that ${P_b} \to \infty $, the $s \to \infty $ and (\ref{b1-1}) can be simplified by keeping the first term of the series, i.e., ${a = b = 0}$. As a consequence, (\ref{b1-1}) is recast as
\begin{small}
\begin{align}\label{b1-2}
\mathcal{L}\left( {{f_{\left| {h_{\varphi,m}h_{b,m}} \right|}}\left( x \right)} \right)\left( s \right){ = _2}{F_1}\left( {2,\frac{1}{2};\frac{5}{2};1} \right)\frac{{\Xi _{\varphi ,\infty }^{pSIC}}}{{{s^2}}}.
\end{align}
\end{small}Since ${{{{\mathbf{\hat H}}}_{b\varphi }}} = \sum\nolimits_{m = 1}^M {\left| {h_{\varphi,m}h_{b,m}} \right|} $, the PDF of ${{{{\mathbf{\hat H}}}_{b\varphi }}}$ can be attained by convolving the ${{f_{\left| {h_{\varphi,m}h_{b,m}} \right|}}\left( x \right)}$, $m = 1,2, \cdots ,M$. Considering the convolution theorem, we have
\begin{small}
\begin{align}\label{b2}
\mathcal{L}\left( {{f_{{{{\mathbf{\hat H}}}_{b\varphi }}}}\left( x \right)} \right)\left( s \right) =& {\left( {\mathcal{L}\left( {{f_{\left| {h_{\varphi,m}h_{b,m}} \right|}}\left( x \right)} \right)\left( s \right)} \right)^M} \notag \\ =& {\left( {_2{F_1}\left( {2,\frac{1}{2};\frac{5}{2};1} \right)\frac{{\Xi _{\varphi ,\infty }^{pSIC}}}{{{s^2}}}} \right)^M}.
\end{align}
\end{small}By taking the inverse Laplace transform of the above equation, we can further obtain the PDF of the cascaded channels ${{{{\mathbf{\hat H}}}_{b\varphi }}}$ given as follows
\begin{small}
\begin{align}\label{b3}
{f_{{{{\mathbf{\hat H}}}_{b\varphi }}}}\left( x \right) =& {\mathcal{L}^{ - 1}}\left( {{{\left( {_2{F_1}\left( {2,\frac{1}{2};\frac{5}{2};1} \right)\frac{{\Xi _{\varphi ,\infty }^{pSIC}}}{{{s^2}}}} \right)}^M}} \right)\notag \\  =& \frac{{{{\left( {16\left( {1 + {\kappa _{b}}} \right)\left( {1 + {\kappa _{\varphi }}} \right)x} \right)}^M}}}{{\left( {2M} \right)!{{\left( {3{e^{{\kappa _{b}} + {\kappa _{\varphi }}}}} \right)}^M}}}.
\end{align}
\end{small}On this basis, the PDF of ${{{\left| {{{{\mathbf{\hat H}}}_{b\varphi }}} \right|}^2}}$ can be derived referring to the subsequence relationship
\begin{small}
\begin{align}\label{b4}
{f_{{{\left| {{{{\mathbf{\hat H}}}_{b\varphi }}} \right|}^2}}}\left( x \right) = \frac{1}{{2\sqrt x }}\left( {{f_{{{{\mathbf{\hat H}}}_{b\varphi }}}}\left( {\sqrt x } \right) + {f_{{{{\mathbf{\hat H}}}_{b\varphi }}}}\left( { - \sqrt x } \right)} \right).
\end{align}
\end{small}The corresponding CDF of ${{{\left| {{{{\mathbf{\hat H}}}_{b\varphi }}} \right|}^2}}$ can be acquired by imposing derivation operation on (\ref{b4}). With following the similar procedure of (\ref{a4})-(\ref{a7}) in Appendix A and replacing the original ${F_X}\left(  \cdot  \right)$ in (\ref{a5}), the asymptotic SOP expression of $u_r$ with perfect SIC is derived as shown in (\ref{asy sop ee r psic}). The proof is completed.

\section*{Appendix~C: Proof of Theorem \ref{theorem3} } \label{AppendixC}
\renewcommand{\theequation}{C.\arabic{equation}}
\setcounter{equation}{0}
The proof starts with defining the secrecy outage event of $u_r$ in the internal wiretapping scenario, which is shown as
$P_{ie,t}^{ipSIC} = \Pr \left( {\left(\log \left( {1 + \gamma _{t,t}^{ipSIC}} \right) - \log \left( {1 + \gamma _{ie,t}^{ipSIC}} \right)\right) ^ +  < {R_t}} \right)$. Upon substituting (\ref{SINR t decode t}) and (\ref{SINR r decode t}) into it 
and supposing that $X = {\left| {{{{\mathbf{\hat H}}}_{bt}}} \right|^2}$, $V = {\left| {{{{\mathbf{\hat H}}}_{br}}} \right|^2}$, $Y = d_{t}^\alpha $ and $Z = d_{r}^\alpha $, $P_{ie,t}^{ipSIC}$ can be further rewritten as
\begin{small}
\begin{align}\label{c2}
P_{ie,t}^{ipSIC} =& \frac{l_r^{ - {k_r}}}{{2\Gamma \left( {{k_t}} \right)\Gamma \left( {{k_r}} \right)}}\int_0^{R_d^\alpha } {\int_0^{R_d^\alpha } {{f_Z}\left( v \right){f_Y}\left( v \right)dydz} } \notag \\ &\times \int_0^\infty  {{v^{\frac{{{k_r}}}{2} - 1}}{e^{ - \frac{{\sqrt v }}{{{l_r}}}}}\gamma \left( {{k_t},\frac{{\sqrt {{\Delta _2}\left( {z,v,y} \right)} }}{{{l_t}}}} \right)} dv,
\end{align}
\end{small}where ${\Delta _2}\left( {z,v,y} \right) = \hat \Psi \left( {z,v} \right)\left( {{\mu _{rt3}} + \left( {{\mu _{rt2}} + \sigma _n^2} \right)y} \right)/{\mu _{rt1}}$ and $\hat \Psi \left( {z,v} \right) = {2^{{R_t}}}\left( {1 + \frac{{{\varepsilon _{rt1}}v}}{{{\varepsilon _{rt3}} + \left( {{\varepsilon _{rt2}} + \sigma _e^2} \right)z}}} \right) - 1$. Due to the lack of accurate approximation method, the integral in (\ref{c3}) is hard to be tackled directly. For the purpose of tractability, the expectation operation is adopted to approximate the statistical properties of the cascaded internal wiretapping channels, which is shown as

\begin{small}
\begin{align}\label{c3}
\mathbb{E}\left( {{{\left| {{{{\mathbf{\hat H}}}_{br}}} \right|}^2}} \right) =& \mathbb{D}\left( {\left| {{{{\mathbf{\hat H}}}_{br}}} \right|} \right) + {\left( {\mathbb{E}\left( {\left| {{{{\mathbf{\hat H}}}_{br}}} \right|} \right)} \right)^2} \notag \\  =& M\mathbb{D}\left( {\left| {h_{r,m}h_{b,m}} \right|} \right) + {\left( {M\mathbb{E}\left( {\left| {h_{r,m}h_{b,m}} \right|} \right)} \right)^2},
\end{align}
\end{small}where ${\mathbb{E}\left( {\left| {h_{r,m}h_{b,m}} \right|} \right)}$ and $\mathbb{D}\left( {\left| {h_{r,m}h_{b,m}} \right|} \right)$ are given in (\ref{mean single Rician channel}) and (\ref{var single Rician channel}), respectively. They can also be expressed in the form of Bessel functions as follows:
\begin{small}
\begin{align}\label{c4}
\mathbb{E}\left( {\left| {h_{r,m}h_{b,m}} \right|} \right) = \frac{{\pi {e^{ - \frac{1}{2}\left( {{\kappa _{b}} + {\kappa _{r}}} \right)}}}}{{4\sqrt {\left( {{\kappa _{b}} + 1} \right)\left( {{\kappa _{r}} + 1} \right)} }}{\mathcal{I}_{b}}{\mathcal{I}_{r}},
\end{align}
\end{small}
and
\begin{small}
\begin{align}\label{c5}
\mathbb{D}\left( {\left| {h_{r,m}h_{b,m}} \right|} \right) = 1 - \frac{{{\pi ^2}{e^{ - \left( {{\kappa _{b}} + {\kappa _{r}}} \right)}}\mathcal{I}_{b}^2\mathcal{I}_{r}^2}}{{16\left( {{\kappa _{b}} + 1} \right)\left( {{\kappa _{r}} + 1} \right)}},
\end{align}
\end{small}where ${\mathcal{I}_{b}} = \left( {{\kappa _{b}} + 1} \right){I_0}\left( {\frac{{{\kappa _{b}}}}{2}} \right) + {\kappa _{b}}{I_1}\left( {\frac{{{\kappa _{b}}}}{2}} \right)$ and ${\mathcal{I}_{r}} = \left( {{\kappa _{r}} + 1} \right){I_0}\left( {\frac{{{\kappa _{r}}}}{2}} \right) + {\kappa _{r}}{I_1}\left( {\frac{{{\kappa _{r}}}}{2}} \right)$ and ${I_v}\left(  \cdot  \right)$ indicates the modified \emph{v}-order Bessel function of the first kind \cite[Eq. (8.431)]{gradvstejn2000table}. Hence, the expectation of ${{{\left| {{{{\mathbf{\hat H}}}_{br}}} \right|}^2}}$ is given by
\begin{small}
\begin{align}\label{c6}
{\Omega _{br}} =& M\left( {1 - \frac{{{\pi ^2}{e^{ - \left( {{\kappa _{b}} + {\kappa _{r}}} \right)}}\mathcal{I}_{b}^2\mathcal{I}_{r}^2}}{{16\left( {{\kappa _{b}} + 1} \right)\left( {{\kappa _{r}} + 1} \right)}}} \right) \notag \\ &+ {\left( {\frac{{M\pi {e^{ - \frac{1}{2}\left( {{\kappa _{b}} + {\kappa _{r}}} \right)}}}}{{4\sqrt {\left( {{\kappa _{b}} + 1} \right)\left( {{\kappa _{r}} + 1} \right)} }}{\mathcal{I}_{b}}{\mathcal{I}_{r}}} \right)^2}.
\end{align}
\end{small}By substituting (\ref{c6}) into (\ref{c2}), we have $P_{ie,t}^{ipSIC}\left( {{R_t}} \right) = \int_0^{R_d^\alpha } {\int_0^{R_d^\alpha } {{f_Z}\left( v \right){f_Y}\left( v \right){F_X}\left( {\frac{{\hat \Psi \left( z \right)\left( {{\mu _{rt3}} + \left( {{\mu _{rt2}} + \sigma _n^2} \right)y} \right)}}{{{\mu _{rt1}}}}} \right)dydz} } $ and $\hat \Psi \left( z \right) = {2^{{R_t}}}\left( {1 + \frac{{{\varepsilon _{rt1}}{\Omega _{br}}}}{{{\varepsilon _{rt3}} + \left( {{\varepsilon _{rt2}} + \sigma _e^2} \right)z}}} \right) - 1$. Referring to the procedure illustrated in (\ref{a3})-(\ref{a7}) and applying some simple calculations, we can obtain (\ref{sop ie t ipsic}). The proof is completed.

\bibliographystyle{IEEEtran}
\bibliography{MF_RIS_PLS}

\begin{thebibliography}{10}
\providecommand{\url}[1]{#1}
\csname url@samestyle\endcsname
\providecommand{\newblock}{\relax}
\providecommand{\bibinfo}[2]{#2}
\providecommand{\BIBentrySTDinterwordspacing}{\spaceskip=0pt\relax}
\providecommand{\BIBentryALTinterwordstretchfactor}{4}
\providecommand{\BIBentryALTinterwordspacing}{\spaceskip=\fontdimen2\font plus
\BIBentryALTinterwordstretchfactor\fontdimen3\font minus
  \fontdimen4\font\relax}
\providecommand{\BIBforeignlanguage}[2]{{%
\expandafter\ifx\csname l@#1\endcsname\relax
\typeout{** WARNING: IEEEtran.bst: No hyphenation pattern has been}%
\typeout{** loaded for the language `#1'. Using the pattern for}%
\typeout{** the default language instead.}%
\else
\language=\csname l@#1\endcsname
\fi
#2}}
\providecommand{\BIBdecl}{\relax}
\BIBdecl

\bibitem{Zhengquan20196G}
Z.~Zhang, Y.~Xiao, Z.~Ma, M.~Xiao, Z.~Ding, X.~Lei, G.~K. Karagiannidis, and
  P.~Fan, ``6{G} wireless networks: Vision, requirements, architecture, and key
  technologies,'' \emph{{IEEE} Veh. Technol. Mag.}, vol.~14, no.~3, pp. 28--41,
  Sep. 2019.

\bibitem{you2021towards}
X.~You, C.-X. Wang, J.~Huang, X.~Gao, Z.~Zhang, M.~Wang, Y.~Huang, C.~Zhang,
  Y.~Jiang, J.~Wang \emph{et~al.}, ``Towards 6{G} wireless communication
  networks: Vision, enabling technologies, and new paradigm shifts,''
  \emph{Sci. China Inf. Sci.}, vol.~64, no.~1, pp. 1--74, 2021.

\bibitem{Saad2020vision6G}
W.~Saad, M.~Bennis, and M.~Chen, ``A vision of 6{G} wireless systems:
  Applications, trends, technologies, and open research problems,''
  \emph{{IEEE} Netw.}, vol.~34, no.~3, pp. 134--142, May 2020.

\bibitem{2020QingqingMag}
Q.~Wu and R.~Zhang, ``Towards smart and reconfigurable environment: Intelligent
  reflecting surface aided wireless network,'' \emph{{IEEE} Commun. Mag.},
  vol.~58, no.~1, pp. 106--112, Jan. 2020.

\bibitem{ShiminGong2020RIS}
S.~Gong, X.~Lu, D.~T. Hoang, D.~Niyato, L.~Shu, D.~I. Kim, and Y.-C. Liang,
  ``Toward smart wireless communications via intelligent reflecting surfaces: A
  contemporary survey,'' \emph{{IEEE} Commun. Surveys Tutorials}, vol.~22,
  no.~4, pp. 2283--2314, Jun 2020.

\bibitem{2021QingqingRISTutorial}
Q.~Wu, S.~Zhang, B.~Zheng, C.~You, and R.~Zhang, ``Intelligent reflecting
  surface-aided wireless communications: A tutorial,'' \emph{{IEEE} Trans.
  Commun.}, vol.~69, no.~5, pp. 3313--3351, May 2021.

\bibitem{2020BeixiongRISOFDM}
Y.~Yang, B.~Zheng, S.~Zhang, and R.~Zhang, ``Intelligent reflecting surface
  meets {OFDM}: Protocol design and rate maximization,'' \emph{{IEEE} Trans.
  Commun.}, vol.~68, no.~7, pp. 4522--4535, Jul. 2020.

\bibitem{LiangYang2020RISCoverage}
L.~Yang, Y.~Yang, M.~O. Hasna, and M.-S. Alouini, ``Coverage, probability of
  {SNR} gain, and {DOR} analysis of {RIS}-aided communication systems,''
  \emph{{IEEE} Wireless Commun. Lett.}, vol.~9, no.~8, pp. 1268--1272, Aug.
  2020.

\bibitem{Jingzhi2020RISSensing}
J.~Hu, H.~Zhang, B.~Di, L.~Li, K.~Bian, L.~Song, Y.~Li, Z.~Han, and H.~V. Poor,
  ``Reconfigurable intelligent surface based {RF} sensing: Design,
  optimization, and implementation,'' \emph{{IEEE} J. Sel. Areas Commun.},
  vol.~38, no.~11, pp. 2700--2716, Nov. 2020.

\bibitem{Xinyi2022RISSensing}
X.~Wang, Z.~Fei, J.~Huang, and H.~Yu, ``Joint waveform and discrete phase shift
  design for {RIS}-assisted integrated sensing and communication system under
  {C}ramer-{R}ao bound constraint,'' \emph{{IEEE} Trans. Veh. Technol.},
  vol.~71, no.~1, pp. 1004--1009, Jan. 2022.

\bibitem{GuenSun2021RISUAVPLS}
G.~Sun, X.~Tao, N.~Li, and J.~Xu, ``Intelligent reflecting surface and {UAV}
  assisted secrecy communication in millimeter-wave networks,'' \emph{{IEEE}
  Trans. Veh. Technol.}, vol.~70, no.~11, pp. 11\,949--11\,961, Nov. 2021.

\bibitem{Yingjie2023RISPLS}
Y.~Pei, X.~Yue, W.~Yi, Y.~Liu, X.~Li, and Z.~Ding, ``Secrecy outage probability
  analysis for downlink {RIS}-{NOMA} networks with on-off control,''
  \emph{{IEEE} Trans. Veh. Technol.}, vol.~72, no.~9, pp. 11\,772--11\,786,
  Sep. 2023.

\bibitem{Waqas2024RISPLS}
W.~Khalid, M.~A.~U. Rehman, T.~Van~Chien, Z.~Kaleem, H.~Lee, and H.~Yu,
  ``Reconfigurable intelligent surface for physical layer security in
  6{G}-{I}o{T}: Designs, issues, and advances,'' \emph{{IEEE} Internet Things
  J.}, vol.~11, no.~2, pp. 3599--3613, Jan. 2024.

\bibitem{Linglong2020RISPrototyping}
L.~Dai, B.~Wang, M.~Wang, X.~Yang, J.~Tan, S.~Bi, S.~Xu, F.~Yang, Z.~Chen,
  M.~D. Renzo, C.-B. Chae, and L.~Hanzo, ``Reconfigurable intelligent
  surface-based wireless communications: Antenna design, prototyping, and
  experimental results,'' \emph{{IEEE} Access}, vol.~8, pp. 45\,913--45\,923,
  Mar. 2020.

\bibitem{Yuanwei2021STAR}
Y.~Liu, X.~Mu, J.~Xu, R.~Schober, Y.~Hao, H.~V. Poor, and L.~Hanzo, ``{STAR}:
  Simultaneous transmission and reflection for 360° coverage by intelligent
  surfaces,'' \emph{{IEEE} Wireless Commun.}, vol.~28, no.~6, pp. 102--109,
  Dec. 2021.

\bibitem{JiaqiXu2021STAR}
J.~Xu, Y.~Liu, X.~Mu, and O.~A. Dobre, ``{STAR}-{RIS}s: Simultaneous
  transmitting and reflecting reconfigurable intelligent surfaces,''
  \emph{{IEEE} Commun. Lett.}, vol.~25, no.~9, pp. 3134--3138, Sep. 2021.

\bibitem{ChaoZhang2022STARNOMA}
C.~Zhang, W.~Yi, Y.~Liu, Z.~Ding, and L.~Song, ``{STAR}-{IOS} aided {NOMA}
  networks: Channel model approximation and performance analysis,''
  \emph{{IEEE} Trans. Wireless Commun.}, vol.~21, no.~9, pp. 6861--6876, Sep.
  2022.

\bibitem{Papazafeiropoulos2022STARMIMO}
A.~Papazafeiropoulos, Z.~Abdullah, P.~Kourtessis, S.~Kisseleff, and
  I.~Krikidis, ``Coverage probability of {STAR}-{RIS}-assisted massive {MIMO}
  systems with correlation and phase errors,'' \emph{{IEEE} Wireless Commun.
  Lett.}, vol.~11, no.~8, pp. 1738--1742, Aug. 2022.

\bibitem{salem2023star}
A.~Salem, K.-K. Wong, C.-B. Chae, and Y.~Zhang, ``{STAR}-{RIS} assisted
  full-duplex communication networks,'' \emph{arXiv preprint arXiv:2309.15037},
  2023.

\bibitem{Hehao2021STARPLS}
H.~Niu, Z.~Chu, F.~Zhou, and Z.~Zhu, ``Simultaneous transmission and reflection
  reconfigurable intelligent surface assisted secrecy {MISO} networks,''
  \emph{{IEEE} Commun. Lett.}, vol.~25, no.~11, pp. 3498--3502, Nov. 2021.

\bibitem{Guojie2023STARActiveEve}
G.~Hu, Z.~Li, J.~Si, K.~Xu, Y.~Cai, D.~Xu, and N.~Al-Dhahir, ``Analysis and
  optimization of {STAR}-{RIS}-assisted proactive eavesdropping with
  statistical {CSI},'' \emph{{IEEE} Trans. Veh. Technol.}, vol.~72, no.~5, pp.
  6850--6855, May 2023.

\bibitem{Manzoor2023STAR}
M.~Ahmed, A.~Wahid, S.~S. Laique, W.~U. Khan, A.~Ihsan, F.~Xu, S.~Chatzinotas,
  and Z.~Han, ``A survey on {STAR}-{RIS}: Use cases, recent advances, and
  future research challenges,'' \emph{{IEEE} Internet Things J.}, vol.~10,
  no.~16, pp. 14\,689--14\,711, Aug. 2023.

\bibitem{2022XinyueNGMA}
X.~Pei, Y.~Chen, M.~Wen, H.~Yu, E.~Panayirci, and H.~V. Poor, ``Next-generation
  multiple access based on {NOMA} with power level modulation,'' \emph{{IEEE}
  J. Sel. Areas Commun.}, vol.~40, no.~4, pp. 1072--1083, Apr. 2022.

\bibitem{Boqun2022STARNOMA}
B.~Zhao, C.~Zhang, W.~Yi, and Y.~Liu, ``Ergodic rate analysis of {STAR}-{RIS}
  aided {NOMA} systems,'' \emph{{IEEE} Commun. Lett.}, vol.~26, no.~10, pp.
  2297--2301, Apr. 2022.

\bibitem{Xinwei2023STARRIS}
X.~Yue, J.~Xie, Y.~Liu, Z.~Han, R.~Liu, and Z.~Ding, ``Simultaneously
  transmitting and reflecting reconfigurable intelligent surface assisted
  {NOMA} networks,'' \emph{{IEEE} Trans. Wireless Commun.}, vol.~22, no.~1, pp.
  189--204, Jan. 2023.

\bibitem{Farjam2023STARNOMA}
F.~Karim, S.~K. Singh, K.~Singh, S.~Prakriya, and M.~F. Flanagan, ``On the
  performance of {STAR}-{RIS}-aided {NOMA} at finite blocklength,''
  \emph{{IEEE} Wireless Commun. Lett.}, vol.~12, no.~5, pp. 868--872, May 2023.

\bibitem{Brian2023SERSTARNOMA}
B.~Y.~D. Rito and K.~H. Li, ``{SER}-effective constellation scaling and
  rotation in {STAR}-{RIS}-assisted uplink {NOMA},'' \emph{{IEEE} Commun.
  Lett.}, vol.~27, no.~9, pp. 2506--2510, Sep. 2023.

\bibitem{lv2024safeguarding}
L.~Lv, D.~Xu, R.~Q. Hu, Y.~Ye, L.~Yang, X.~Lei, X.~Wang, D.~I. Kim, and
  A.~Nallanathan, ``Safeguarding next-generation multiple access using physical
  layer security techniques: A tutorial,'' \emph{Proc. {IEEE}}, to appear in
  2024.

\bibitem{Xiangbin2023STARNOMAPLS}
X.~Yu, K.~Shen, and X.~Dang, ``Secure performance of {STAR}-{RIS} aided {NOMA}
  system with imperfect {SIC},'' \emph{{IEEE} Wireless Commun. Lett.}, vol.~12,
  no.~12, pp. 2023--2027, Dec. 2023.

\bibitem{li2022enhancing}
X.~Li, Y.~Zheng, M.~Zeng, Y.~Liu, and O.~A. Dobre, ``Enhancing secrecy
  performance for {STAR}-{RIS} {NOMA} networks,'' \emph{{IEEE} Trans. Veh.
  Technol.}, vol.~72, no.~2, pp. 2684--2688, Feb. 2022.

\bibitem{YiHan2022STARNOMAPLS}
Y.~Han, N.~Li, Y.~Liu, T.~Zhang, and X.~Tao, ``Artificial noise aided secure
  {NOMA} communications in {STAR}-{RIS} networks,'' \emph{{IEEE} Wireless
  Commun. Lett.}, vol.~11, no.~6, pp. 1191--1195, Jun. 2022.

\bibitem{HuiHan2023STARNOMAPLS}
H.~Han, Y.~Cao, N.~Deng, C.~Xing, N.~Zhao, Y.~Li, and X.~Wang, ``Secure
  transmission for {STAR}-{RIS} aided {NOMA} against internal eavesdropping,''
  \emph{{IEEE} Trans. Veh. Technol.}, vol.~72, no.~11, pp. 15\,068--15\,073,
  Nov. 2023.

\bibitem{ZhengZhang2022STARPLS}
Z.~Zhang, J.~Chen, Y.~Liu, Q.~Wu, B.~He, and L.~Yang, ``On the secrecy design
  of {STAR}-{RIS} assisted uplink {NOMA} networks,'' \emph{{IEEE} Trans.
  Wireless Commun.}, vol.~21, no.~12, pp. 11\,207--11\,221, Dec. 2022.

\bibitem{YanboZhang2023STARNOMAPLS}
Y.~Zhang, Z.~Yang, J.~Cui, P.~Xu, G.~Chen, Y.~Wu, and M.~D. Renzo,
  ``{STAR}-{RIS} assisted secure transmission for downlink multi-carrier {NOMA}
  networks,'' \emph{{IEEE} Trans. Inf. Forensics Security}, vol.~18, pp.
  5788--5803, Sep. 2023.

\bibitem{zhang2021active}
Z.~Zhang, L.~Dai, X.~Chen, C.~Liu, F.~Yang, R.~Schober, and H.~V. Poor,
  ``Active {RIS} vs. passive {RIS}: Which will prevail in 6{G}?'' \emph{{IEEE}
  Trans. Commun.}, vol.~71, no.~3, pp. 1707--1725, Mar. 2023.

\bibitem{WenWang2023MFRIS}
W.~Wang, W.~Ni, and H.~Tian, ``Multi-functional {RIS}-aided wireless
  communications,'' \emph{{IEEE} Internet Things J.}, vol.~10, no.~23, pp.
  21\,133--21\,134, Dec. 2023.

\bibitem{Wanli2024MFRIS}
W.~Ni, A.~Zheng, W.~Wang, D.~Niyato, N.~Al-Dhahir, and M.~Debbah, ``From single
  to multi-functional {RIS}: Architecture, key technologies, challenges, and
  applications,'' \emph{{IEEE} Netw.}, to appear in 2024.

\bibitem{Yue2024ASTRSNOMA}
X.~Yue, J.~Xie, C.~Ouyang, Y.~Liu, X.~Shen, and Z.~Ding, ``Active
  simultaneously transmitting and reflecting surface assisted {NOMA}
  networks,'' \emph{{IEEE} Trans. Wireless Commun.}, vol.~23, no.~8, pp.
  9912--9926, Aug. 2024.

\bibitem{Ailing2023MFRIS}
A.~Zheng, W.~Ni, W.~Wang, H.~Tian, Y.~C. Eldar, and D.~Niyato,
  ``Multi-functional {RIS}: Signal modeling and optimization,'' \emph{{IEEE}
  Trans. Veh. Technol.}, vol.~73, no.~4, pp. 5971--5976, Apr. 2024.

\bibitem{YingjieYan2023MFRIS}
Y.~Yan, Y.~Wang, W.~Ni, and D.~Niyato, ``Joint beamforming design for
  multi-functional {RIS}-aided uplink communications,'' \emph{{IEEE} Commun.
  Lett.}, vol.~27, no.~10, pp. 2697--2701, Oct. 2023.

\bibitem{Ailing2023NextGenMFRIS}
A.~Zheng, W.~Ni, W.~Wang, and H.~Tian, ``Next-generation {RIS}: From single to
  multiple functions,'' \emph{{IEEE} Wireless Commun. Lett.}, vol.~12, no.~12,
  pp. 1988--1992, Dec. 2023.

\bibitem{Anastasios2024ASTAR}
A.~Papazafeiropoulos, H.~Ge, P.~Kourtessis, T.~Ratnarajah, S.~Chatzinotas, and
  S.~Papavassiliou, ``Two-timescale design for active {STAR}-{RIS} aided
  massive {MIMO} systems,'' \emph{{IEEE} Trans. Veh. Technol.}, to appear in
  2024 2024.

\bibitem{ZheZhang2021PLSRIS}
Z.~Zhang, C.~Zhang, C.~Jiang, F.~Jia, J.~Ge, and F.~Gong, ``Improving physical
  layer security for reconfigurable intelligent surface aided {NOMA} 6{G}
  networks,'' \emph{{IEEE} Trans. Veh. Technol.}, vol.~70, no.~5, pp.
  4451--4463, May 2021.

\bibitem{WenWang2023UAVRISPLS}
W.~Wang, W.~Ni, H.~Tian, Y.~C. Eldar, and D.~Niyato, ``{UAV}-mounted
  multi-functional {RIS} for combating eavesdropping in wireless networks,''
  \emph{{IEEE} Wireless Commun. Lett.}, vol.~12, no.~10, pp. 1667--1671, Oct.
  2023.

\bibitem{WenWang2022STARPLS}
W.~Wang, W.~Ni, H.~Tian, Z.~Yang, C.~Huang, and K.-K. Wong, ``Safeguarding
  {NOMA} networks via reconfigurable dual-functional surface under imperfect
  {CSI},'' \emph{{IEEE} J. Sel. Topics Signal Process.}, vol.~16, no.~5, pp.
  950--966, Aug. 2022.

\bibitem{LuLv2022CovertRISNOMA}
L.~Lv, Q.~Wu, Z.~Li, Z.~Ding, N.~Al-Dhahir, and J.~Chen, ``Covert communication
  in intelligent reflecting surface-assisted {NOMA} systems: Design, analysis,
  and optimization,'' \emph{{IEEE} Trans. Wireless Commun.}, vol.~21, no.~3,
  pp. 1735--1750, Mar. 2022.

\bibitem{ZhengZhang2023STARPLS}
Z.~Zhang, Z.~Wang, Y.~Liu, B.~He, L.~Lv, and J.~Chen, ``Security enhancement
  for coupled phase-shift {STAR}-{RIS} networks,'' \emph{{IEEE} Trans. Veh.
  Technol.}, vol.~72, no.~6, pp. 8210--8215, Jun 2023.

\bibitem{Yingjie2023ARISNOMA}
X.~Li, Y.~Pei, X.~Yue, Y.~Liu, and Z.~Ding, ``Secure communication of active
  {RIS} assisted {NOMA} networks,'' \emph{{IEEE} Trans. Wireless Commun.},
  vol.~23, no.~5, pp. 4489--4503, May 2023.

\bibitem{2022LinglongSunARIS}
K.~Liu, Z.~Zhang, L.~Dai, S.~Xu, and F.~Yang, ``Active reconfigurable
  intelligent surface: Fully-connected or sub-connected?'' \emph{{IEEE} Commun.
  Lett.}, vol.~26, no.~1, pp. 167--171, Jan. 2022.

\bibitem{2022CunhuaPanARIS}
K.~Zhi, C.~Pan, H.~Ren, K.~K. Chai, and M.~Elkashlan, ``Active {RIS} versus
  passive {RIS}: Which is superior with the same power budget?'' \emph{{IEEE}
  Commun. Lett.}, vol.~26, no.~5, pp. 1150--1154, May 2022.

\bibitem{NaLi2021IRSNOMAPLS}
N.~Li, M.~Li, Y.~Liu, C.~Yuan, and X.~Tao, ``Intelligent reflecting surface
  assisted {NOMA} with heterogeneous internal secrecy requirements,''
  \emph{{IEEE} Wireless Commun. Lett.}, vol.~10, no.~5, pp. 1103--1107, May
  2021.

\bibitem{gradvstejn2000table}
I.~Grad{\v{s}}tejn and I.~M. Ry{\v{z}}ik, ``Table of integrals, series, and
  products, 6th ed,'' 2000.

\bibitem{simon2002probability}
M.~K. Simon, \emph{Probability distributions involving Gaussian random
  variables: A handbook for engineers and scientists}.\hskip 1em plus 0.5em
  minus 0.4em\relax Springer, 2002.

\bibitem{primak2005stochastic}
S.~Primak, V.~Kontorovich, and V.~Lyandres, \emph{Stochastic methods and their
  applications to communications: stochastic differential equations
  approach}.\hskip 1em plus 0.5em minus 0.4em\relax John Wiley \& Sons, 2005.

\bibitem{hildebrand1987introduction}
F.~B. Hildebrand, \emph{Introduction to numerical analysis}.\hskip 1em plus
  0.5em minus 0.4em\relax Courier Corporation, 1987.

\end{thebibliography}

\end{document}